\documentclass[journal, twocolumn]{IEEEtran}

\usepackage[utf8]{inputenc} 	

\usepackage{amssymb}
\usepackage{bbm} 
\usepackage{graphicx}
\usepackage{color,psfrag}

\usepackage{MnSymbol}

\usepackage[normalem]{ulem}

\usepackage{tikz}
\usepackage{siunitx}
\usepackage{cite} 

\ifCLASSOPTIONcompsoc
\usepackage[caption=false,font=normalsize,labelfont=sf,textfont=sf]{subfig}
\else
\usepackage[caption=false,font=footnotesize]{subfig}
\fi
\usepackage{xcolor}
\newcommand{\tc}[1]{#1}
\newcommand{\tcg}[1]{#1}

\newcommand{\e}[2]{{\mathbb E}_{#1}\left[ #2 \right]}
\newcommand{\s}[2]{{\frac{1}{{#1}}\sum_n^{#1}} {#2}}

\newcommand{\p}{\mathbb P}
\newcommand{\sub}[1]{_{\text{#1}}}

\newcommand{\opc}{\rho\sub{out}}
\newcommand{\pc}{P\sub{Tx,ST,full}}

\newcommand{\preg}{P\sub{Tx,ST,cont}}
\newcommand{\xreg}{x\sub{ST,cont}}
\newcommand{\prcvd}{P\sub{Rx,ST}}
\newcommand{\prcvdsr}{P\sub{Rx,SR}}
\newcommand{\eprcvd}{\hat{P}\sub{Rx,ST}}
\newcommand{\eprcvdsr}{\hat{P}\sub{Rx,SR}}
\newcommand{\yrcvd}{y\sub{ST}}
\newcommand{\ptran}{P\sub{Tx,PR}}
\newcommand{\ptranpt}{P\sub{Tx,PT}}
\newcommand{\xtran}{x\sub{PR}}
\newcommand{\xtranpt}{x\sub{PT}}
\newcommand{\pp}{P\sub{Rx,PR}}
\newcommand{\yp}{y\sub{PR}}
\newcommand{\ys}{y\sub{SR}}
\newcommand{\nap}{w\sub{PR}}
\newcommand{\nas}{w\sub{ST}}
\newcommand{\nasr}{w\sub{SR}}
\newcommand{\ite}{\theta\sub{I}}
\newcommand{\rs}{R\sub{s}}
\newcommand{\trs}{R\sub{s}}

\newcommand{\gpo}{h\sub{PR,ST}}
\newcommand{\gpt}{h\sub{PT,SR}}

\newcommand{\epgpo}{|\hat{h}\sub{PR,ST}|^2}
\newcommand{\epgpt}{|\hat{h}\sub{PR,SR}|^2}
\newcommand{\epgs}{|\hat{h}\sub{ST,SR}|^2}
\newcommand{\gs}{h\sub{ST,SR}}
\newcommand{\pgpo}{|h\sub{PR,ST}|^2}
\newcommand{\pgpt}{|h\sub{PT,SR}|^2}
\newcommand{\egs}{\hat{h}\sub{s}}
\newcommand{\pgs}{|h\sub{ST,SR}|^2}
\newcommand{\bpgpo}{|\bar{h}\sub{PR,ST}|^2}
\newcommand{\bpgpt}{|\bar{h}\sub{PT,SR}|^2}
\newcommand{\bpgs}{|\bar{h}\sub{ST,SR}|^2}

\newcommand{\nps}{\sigma^2}

\newcommand{\fsam}{f\sub{s}}
\newcommand{\ttau}{\tilde{\tau}}
\newcommand{\ca}{\text{C}\sub{s}}
\newcommand{\eca}{\hat{\text{{C}}}\sub{s}}
\newcommand{\mpo}{m\sub{PR,ST}}
\newcommand{\mpt}{m\sub{PT,SR}}
\newcommand{\ms}{m\sub{ST,SR}}

\newcommand{\fprcvd}{F_{\eprcvd}}
\newcommand{\fpgpo}{F_{\pgpo}}
\newcommand{\fpgpt}{F_{\pgpt}}
\newcommand{\fpgs}{F_{\pgs}}
\newcommand{\fprcvdsr}{F_{\eprcvdsr}}

\newcommand{\fgs}{F_{\epgs}}

\newcommand{\fc}{F_{\eca}}

\newcommand{\dc}{f_{\eca}}

\newcommand{\lpo}{\lambda\sub{p,1}}
\newcommand{\lpt}{\lambda\sub{p,2}}
\newcommand{\ls}{\lambda\sub{s}}
\newcommand{\Ks}{\tau\sub{p}{\fsam}}

\newcommand{\apo}{a\sub{p,1}}
\newcommand{\bpo}{b\sub{p,1}}
\newcommand{\apt}{a\sub{p,2}}
\newcommand{\bpt}{b\sub{p,2}}
\newcommand{\as}{a\sub{s}}
\newcommand{\bs}{b\sub{s}}
\newcommand{\tp}{\frac{\tau\sub{p}}{2}}

\DeclareMathOperator*{\maxi}{max}

\DeclareMathOperator*{\ncchi2}{{\mathcal{X}\sp{\prime}}^2}

\newtheorem{theorem}{\tc{Problem}}

\newtheorem{lemma}{Lemma}

\newtheorem{remark}{Remark}
\newtheorem{coro}{Corollary}

\newtheorem{approxi}{Approximation}

\makeatletter
\if@twocolumn
	\newcommand{\figscale}{0.92 \columnwidth}
	\newcommand{\figscalet}{\columnwidth}
	\newcommand{\figscalett}{\columnwidth}
\else
	\newcommand{\figscale}{0.44 \columnwidth}
	\newcommand{\figscalet}{0.46 \columnwidth}
	\newcommand{\figscalett}{0.68 \columnwidth}
\fi
\makeatother

\makeatletter
\if@twocolumn
	\newcommand{\figmara}{0mm}
	\newcommand{\figmarb}{-2mm}
	\newcommand{\figmarc}{0mm}
	\newcommand{\figmard}{0mm} 
	\newcommand{\subfigmar}{4mm}
	\newcommand{\tablespacing}{1.4}
	\newcommand{\tmara}{-0.0cm}
	\newcommand{\tmarb}{-0.0cm}
	\newcommand{\absmara}{-0.0cm}
	\newcommand{\absmarb}{-0.0cm}
	\newcommand{\keymara}{-0.0cm}
	\newcommand{\keymarb}{-0.0cm}
\else
	\newcommand{\figmara}{-4mm}
	\newcommand{\figmarb}{-10mm}
	\newcommand{\figmarc}{-5mm} 
	\newcommand{\figmard}{-3mm} 
	\newcommand{\subfigmar}{2mm}
	\newcommand{\tablespacing}{0.85}
	\newcommand{\tmara}{-0.6cm}
	\newcommand{\tmarb}{-0.4cm}
	\newcommand{\absmara}{-2.0cm}
	\newcommand{\absmarb}{-0.2cm}
	\newcommand{\keymara}{-0.3cm}
	\newcommand{\keymarb}{-0.3cm}
\fi
\makeatother

\usepackage{hyperref}

\allowdisplaybreaks

\begin{document}
\title{On the Performance Analysis of Underlay Cognitive Radio Systems: A Deployment Perspective}
\author{Ankit Kaushik\IEEEauthorrefmark{1}, \IEEEmembership{Student Member, IEEE}, Shree Krishna Sharma\IEEEauthorrefmark{2}, \IEEEmembership{Member, IEEE}, \\ Symeon Chatzinotas\IEEEauthorrefmark{2}, \IEEEmembership{Senior Member, IEEE},  Bj\"orn Ottersten\IEEEauthorrefmark{2}, \IEEEmembership{Fellow, IEEE}, \\ Friedrich K. Jondral\IEEEauthorrefmark{1} \IEEEmembership{Senior Member, IEEE}
\thanks{\IEEEauthorrefmark{1}A. Kaushik and F. K. Jondral are with Communications Engineering Lab, Karlsruhe Institute of Technology (KIT), Germany. Email:{\{ankit.kaushik,friedrich.jondral\}@kit.edu.}}
\thanks{\IEEEauthorrefmark{2}S.K. Sharma, S. Chatzinotas and B. Ottersten are with SnT - securityandtrust.lu, University of Luxembourg, Luxembourg. Email:{\{shree.sharma, symeon.chatzinotas, bjorn.ottersten\}@uni.lu}.}
\thanks{The preliminary analysis of this paper has been presented at ICC 2015 in London, UK \cite{Kaushik15}.}
\thanks{This work was partially supported by the National Research Fund, Luxembourg under the CORE projects ``SeMIGod'' and ``SATSENT''.}
}

\maketitle
\thispagestyle{empty}
\pagestyle{empty}

\vspace{\absmara}
\begin{abstract}
\vspace{\absmarb}
We study the performance of cognitive Underlay System (US) that employ power control mechanism at the Secondary Transmitter (ST) from a deployment perspective. Existing baseline models considered for performance analysis either assume the knowledge of involved channels at the ST or retrieve this information by means of a band manager or a feedback channel, however, such situations rarely exist in practice. Motivated by this fact, we propose a novel approach that incorporates estimation of the involved channels at the ST, in order to characterize the performance of the US in terms of interference power received at the primary receiver and throughput at the secondary receiver (or \textit{secondary throughput}). Moreover, we apply an outage constraint that captures the impact of imperfect channel knowledge, particularly on the uncertain interference. Besides this, we employ a transmit power constraint at the ST to classify the operation of the US in terms of an interference-limited regime and a power-limited regime. In addition, we characterize the expressions of the uncertain interference and the secondary throughput for the case where the involved channels encounter Nakagami-$m$ fading. Finally, we investigate a fundamental tradeoff between the estimation time and the secondary throughput depicting an optimized performance of the US.

\end{abstract}
\vspace{\keymara}
\begin{IEEEkeywords}
\vspace{\keymarb}
Cognitive radio, Underlay system, Channel estimation, Estimation-throughput tradeoff 
\end{IEEEkeywords}

\section{Introduction}
Cognitive Radio (CR) communications is considered as one of the viable solutions that addresses the problem of spectrum scarcity of future wireless networks. Secondary access to the licensed spectrum can be broadly categorized into different CR paradigms, namely, interweave, underlay and overlay systems \cite{Goldsmith09}. Among these, underlay and interweave systems are largely associated with the techniques that are applicable at the physical layer and therefore can be considered feasible for hardware deployment. 
The interweave systems employ spectrum sensing to detect the presence of primary user signals while avoiding harmful interference to the primary system. On the other hand, an Underlay System (US) exploits the interference tolerance capability of the primary systems that allows the secondary users to transmit even in the presence of the primary users. To accomplish this, the US employs techniques such as power control to maintain the interference (power) received at the Primary Receiver (PR) below a specified level, defined as Interference Threshold (IT) \cite{Xing07}. In this paper, we focus on the performance characterization of the US that employs power control at the Secondary Transmitter (ST).  

\subsection{Motivation and Related Work}
\tc{In order to enable shared access to the licensed spectrum, it is essential to characterize the performance of a CR system \tcg{in reference to the primary and the secondary systems.} With regard to the primary system, the performance of a US is characterized in terms of interference received at the PR, which arises due to concurrent data transmission over the same channel by the secondary system.} Recently, power control at the ST has emerged as an effective way of regulating the interference induced by the ST. However, the power control primarily requires the knowledge\footnote{Here, the knowledge refers to the channel state information.} of the \textit{primary interference} channel between the ST and the PR at the ST. The preliminary investigations \cite{Xing07, Ghasemi07, Sura08, Musa09_, Kang09}, considered for the performance evaluation of the US, assume this knowledge to be perfectly known at the ST. Such situations rarely exist in practical implementations. In order to address this, the performance analysis based on imperfect channel knowledge has been dealt extensively in \cite{Musa09, Suraweera10, Kim12, Alou12, Stat12, Alou13, Zhang13, Smith13, Li13, Kerr14, Sharma15}. 

\tc{It is worth noticing that the majority of these works \cite{Musa09, Alou12, Kim12} in reference to the imperfect channel knowledge consider that the channel's knowledge at the ST is obtained from a band manager\footnote{An entity that mediates between the primary and the secondary system.}, an approach proposed in \cite{Peha05}. Whereas \cite{Alou13, Suraweera10} rely on the presence of a feedback link from the PR to the ST \cite{Zhang08}. The fact is, the feasibility of the band manager or the feedback link across two different systems is unrealistic from a practical standpoint. In addition, due to latency, the channel knowledge obtained while implementing these approaches may be outdated, as considered in \cite{Suraweera10, Kim12, Alou12, Alou13}. \tcg{Besides, for the existence of the feedback link, the demodulation of the secondary user signals at the PR and a resource (time) allocation explicitly for communicating the channel knowledge impose an additional overhead for the primary system. These issues render the hardware implementation of the US in reference to the aforementioned approaches challenging.} In contrast to these approaches, we propose a novel strategy, according to which the channel estimation is employed directly at the secondary system. Thus, by avoiding the realization of the band manager or the feedback link and the issues related to it, this paper outlines the key aspects that facilitate the hardware deployment of the US.} 

\tc{Along with the performance of the primary system, the achievable data rate at the Secondary Receiver (SR) for the link between the ST and the SR contributes significantly to the overall performance of the US \cite{Musa09, Kang09, Suraweera10, Smith13, Zhang13, Alou12, Alou13}. As a matter of fact, the knowledge of the data rate at the ST can be utilized for guaranteeing a certain quality of service, \tcg{which enables us to visualize potential applications or prominent use cases for the CR system.} 
}In order to characterize the data rate, the ST (along with the primary interference channel, which is associated with power control mechanism) requires the knowledge of \textit{access} channel between the ST and the SR, and \textit{secondary interference} channel between the Primary Transmitter (PT) and the SR. \tc{Despite these facts, the performance characterization of the US's data rate in reference to the estimation of the access and the secondary interference channels has not been considered in \cite{Musa09, Suraweera10, Kim12, Alou12, Alou13, Zhang13} or only marginally in \cite{Stat12, Smith13, Li13}.} 

\tc{From a deployment perspective, it is worthy to understand that the interference channels are representative of the channels that exist between different (primary and secondary) systems. This signifies that in order to carry out channel estimation based on the conventional techniques such as pilot-based channel estimation, which is mainly employed in the previous works, a preliminary processing of the primary user signal is necessary. The existence of multiple wireless standards and their complexity forbid the deployment of a dedicated circuitry corresponding to each primary user signals \cite{Cabric06}. 
\tcg{In this regard, in order to facilitate hardware deployment of the US, it is necessary to select the estimation techniques such that the complexity and the versatility (to the unknown primary user signals) requirements are satisfied.} 
In this paper, similar to \cite{Kaushik16}, we address this critical problem by employing a received power-based estimation at the ST and the SR for the interference channels. In contrast to the interweave scenario considered in \cite{Kaushik16}, we investigate an underlay scenario herein. 
Recently, a successful deployment of the received power-based channel estimation at the ST in context to the US has been studied in \cite{Kaushik16_CC}.} 

\tcg{In addition, \cite{Musa09, Suraweera10, Kim12, Alou12, Stat12, Alou13, Zhang13, Smith13, Li13, Kerr14} consider that the PR employs pilot-based channel estimation for the channel PR-ST, which is possible only if the PR is willing: (i) to allocate time resources, (ii) to assign a dedicated circuitry for demodulating the secondary user signals and (iii) to establish a feedback link to the ST, thereby challenging the hardware implementation of the US. 
In contrast, for the proposed received power-based estimation, a certain time needs to be allocated by the secondary user for channel estimation that affects the \textit{secondary throughput}.} 
Since the aspect concerning the time allocation for the channel estimation has not been taken into account in any of the previous investigations related to the cognitive US \cite{Musa09, Suraweera10, Kim12, Alou12, Stat12, Alou13, Zhang13, Smith13, Li13, Kerr14}, the performance of the US in terms of the secondary throughput is overestimated. \tcg{Moreover, the imperfect knowledge of the primary interference channel leads to an uncertainty in the interference at the PR, which in certain cases may exceed the IT.} Under such conditions, the conventional constraint imposed in \cite{Xing07, Ghasemi07, Kang09, Musa09_} is strictly violated. As a result, this \textit{uncertain interference} originated from imperfect channel knowledge may seriously degrade the performance of the primary systems. \tc{In order to tackle this issue, we propose to employ an outage constraint that regulates the uncertain interference caused at the PR.} 

\tcg{Besides, through analysis (performed later in Section \ref{sec:th_ana}), it is revealed that the uncertain interference is associated with the estimation time and the controlled power. 
In this context, the estimation time is indirectly associated with the secondary throughput through the controlled power, signifying the influence of the imperfect channel knowledge. On the other side, the time allocation directly affects the secondary throughput.} 
  In this paper, we examine this relationship between the estimation time and the secondary throughput while constraining the uncertain interference below a desired level. \tcg{Although the previous studies have considered channel estimation, the effect of the imperfect channel knowledge in terms of the time allocation and the uncertain interference in context to the US is still underdeveloped.}

\subsection{Contributions}
In this paper, we provide the following contributions: 
\subsubsection{Analytical Framework}
The main contribution of this paper is to derive an analytical framework for underlay CR systems that employ a power control mechanism and incorporate the estimation of the following interacting channels: (i) primary interference channel between the ST and the PR, (ii) secondary interference channel between the PT and the SR, and (iii) access channel between the ST and the SR. \tc{In contrast to the existing works that demand the presence of a band manager or a feedback link in order to retrieve channel knowledge, we propose to employ channel estimation at the secondary system. In order to facilitate the deployment of the \tcg{US}, we propose to employ received power-based channel estimation, specially for the interference channels so that low complexity and versatility requirements to estimate primary user signals is accomplished. Clearly, the channel estimation is detrimental (in terms of the time allocation and the uncertain interference) to the performance of the \tcg{US}, leading to performance degradation. By comparing its performance with the ideal scenario (with perfect channel knowledge), we study the performance degradation caused due to the imperfect channel knowledge.} 

Besides, we characterize the variations due to the imperfect channel knowledge in the performance parameters, which include interference at the PR and throughput at the SR in terms of their cumulative distribution functions (cdfs) pertaining to the deterministic \tc{(not random)} and the random behavior (channel fading) of the interacting channels. Particularly, these variations lead to uncertainty in the interference that may seriously disrupt the operation of the primary system. To regulate this uncertain interference below a tolerable limit, we propose to employ an outage constraint over the uncertain interference. 
\subsubsection{Interference-limited and Power-limited Regimes}
\tc{The power control at the ST depends on the received signal from the PR to noise power ratio at the ST over the link between the PR and the ST, which characterizes the quality of the primary interference channel. In this paper, we characterize the controlled power in terms of the estimation time and the signal to noise ratio such that the outage constraint is satisfied. In practice, the controlled power is limited by the maximum transmit power. Due to this limitation, good channel conditions (which correspond to a low signal to noise power) do not translate into performance gains for the \tcg{US}. We study this behavior of the \tcg{US} in terms of the performance bound, which is illustrated as a relation between the received signal to noise ratio and the estimation time. As depicted later in \figurename~\ref{fig:or}, based on this performance bound, we classify the operation of the US as the interference-limited and the power-limited regimes.} 
\subsubsection{Estimation-throughput Tradeoff}
\tc{Besides, we propose a successful incorporation of the time allocated for the channel estimation in the secondary system's frame structure. 
The time resources dedicated to channel estimation cause a linear decrease in the secondary throughput. 
Therefore, a low estimation time increases the secondary throughput, since less time is allocated for the channel estimation. 
On the other side, its low value increases the uncertain interference, thus, requires a severe power control that ultimately reduces the secondary throughput. 
We study the association of the estimation time in reference to the time allocation and the controlled power to derive a fundamental tradeoff between the estimation time and the secondary throughput such that the uncertain interference is kept below a desired level. 
We employ this tradeoff to derive a suitable estimation time that achieves a maximum secondary throughput for the US. In other words, the considered tradeoff signifies the fact that the performance degradation in terms of the secondary throughput can be effectively controlled through an appropriate selection of the estimation time.} 
\subsubsection{Estimation-dominant and Channel-dominant Regimes}
For the random channel, we classify the variations in the interference arising due to channel estimation and channel fading as an estimation-dominant regime and a channel-dominant regime, respectively. Based on this analysis, it is revealed that a suitable selection of the estimation time leads to the performance (in terms of the secondary throughput) closer to the one predicted by the existing models that consider the perfect channel knowledge of the interacting channels. 

\subsection{Organization}
The subsequent sections of the paper are organized as follows: Section \ref{sec:sys_mod} presents the system model that describes the deployment scenario, the medium access and the signal model. It further presents the problem description and the proposed approach. Section \ref{sec:th_ana} characterizes the cdfs of the performance parameters and establishes the estimation-throughput tradeoff. Section \ref{sec:num_ana} analyzes the numerical results based on the obtained expressions. Finally, Section \ref{sec:conc} concludes the paper. Table \ref{tb:tb1} lists the definitions of acronyms and important mathematical notations used throughout the paper.

\begin{table}
\vspace{\tmara}
\renewcommand{\arraystretch}{\tablespacing}
\caption{Definitions of Acronyms and Notations used}
\vspace{\tmarb}
\label{tb:tb1}
\centering
\scriptsize{
\begin{tabular}{p{0.25\columnwidth}||p{0.7\columnwidth}}
\hline
\bfseries Acronyms and Notations & \bfseries Definitions \\
\hline\hline
CR & Cognitive Radio\\ \hline
IM, EM & Ideal Model, Estimation Model \\ \hline
US & Underlay System \\ \hline
PT, PR, ST, SR & Primary Transmitter, Primary Receiver, Secondary Transmitter, Secondary Receiver \\ \hline
$\fsam$ & Sampling frequency\\ \hline
$\tau, \tau\sub{p}, T$ & Estimation time interval (received power-based estimation), Estimation time interval (pilot-based channel estimation), Frame duration\\ \hline
$\opc$ & Outage probability constraint \\ \hline
$\pc$ & Maximum transmit power or transmit power constraint at ST \\ \hline
$\preg$ & Control power at ST \\ \hline
$\gpo, \gpt, \gs$ & Channel coefficient for the link PR-ST, PT-SR, ST-SR \\ \hline
$\gamma$ & Signal to noise ratio for the link PR-ST at ST \\ \hline
$\rs, \ca$ & Throughput at SR, Data rate at SR \\ \hline
$F_{(\cdot)}$ & Cumulative distribution function of random variable $(\cdot)$\\ \hline
$f_{(\cdot)}$ & Probability density function of random variable $(\cdot)$\\ \hline
$\hat{(\cdot)}$ & Estimated value of ($\cdot$)\\ \hline
$\tilde{(\cdot)}$ & Suitable value of the parameter ($\cdot$) that achieves maximum performance \\ \hline
$\mathbb E_{(\cdot)}$ & Expectation with respect to ($\cdot$) \\ \hline
$\p$ & Probability measure \\ \hline
$\tcg{\ptranpt, \ptran}$ & \tcg{Transmit power at PT and PR}\\ \hline
$\prcvd, \prcvdsr, \pp$ & Power received at ST from PR, interference power received at SR from PT, interference power received at PR from ST\\ \hline
$\nps$ & noise variance for primary and secondary systems\\ \hline
\end{tabular}
}
\end{table}

\section{System Model} \label{sec:sys_mod}
\begin{figure}[!t]
\centering
\includegraphics[width = \figscalet]{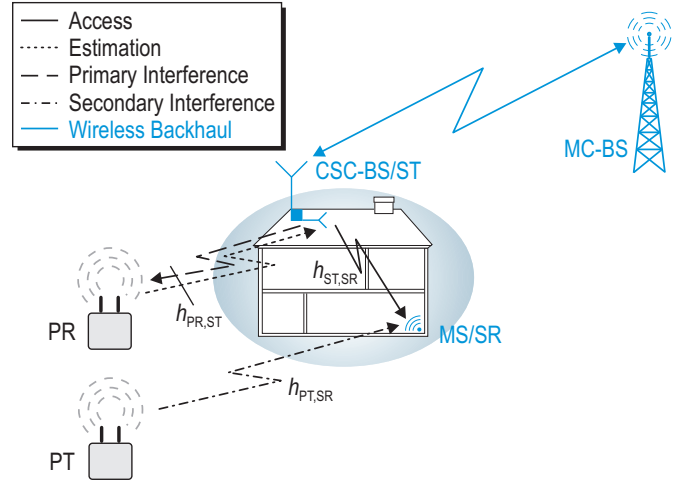}
\vspace{\figmarc}
\caption{\tcg{A cognitive small cell scenario demonstrating: (i) the underlay paradigm, (ii) the associated network elements, which constitute Cognitive Small Cell-Base Station/Secondary Transmitter (CSC-BS/ST), Mobile Station/Secondary Receiver (MS/SR), Macro Cell-Base Station (MC-BS) and Primary Transmitter (PT), (iii) the interacting channels: primary interference channel, secondary interference channel and access channel.}}
\label{fig:scenario}
\vspace{\figmarb}
\end{figure}

\subsection{Underlay Scenario and Medium Access}

The Cognitive Small Cell (CSC), a CR application, characterizes a small cell deployment that fulfills the spectral requirements of the Mobile Stations (MSs) operating indoor,  \figurename~\ref{fig:scenario}.
For the disposition of the CSC in the network, the following key elements are essential: a CSC-Base Station (CSC-BS), a Macro Cell-Base Station (MC-BS) and an MS \cite{Kaushik16}.
Considering the fact that the power control is employed at the CSC-BS, the CSC-BS and the MS represent the ST and the SR, respectively. In order to acquire the knowledge concerning the primary interference channel, the ST listens to the transmissions from the PR. In this work, we consider those primary systems where the PR performs transmissions interchangeably over time (time division duplexing TDD and half-duplex frequency division duplexing FDD) or frequency (full-duplex FDD) with the PT. These transmissions can occur over the same band (TDD) or over separate bands (half-duplex and full-duplex FDD). 
In cellular networks, these duplexing modes are effectively deployed in the Long Term Evolution (LTE) standard \cite{LTE09}. The ST follows these duplexing modes to exploit channel reciprocity principle and determine the interference received at the PR, thus, controls its power for transmitting signals over the access channel such that it satisfies the outage constraint by operating at the IT. Particularly for the half-duplex and the full duplex FDD, it is assumed that the coherence bandwidth is large as compared to the frequency separation between the estimation channel and the band of interest. 

\begin{figure}[!t]
\vspace{\figmara}
\centering
\includegraphics[width = \figscalett]{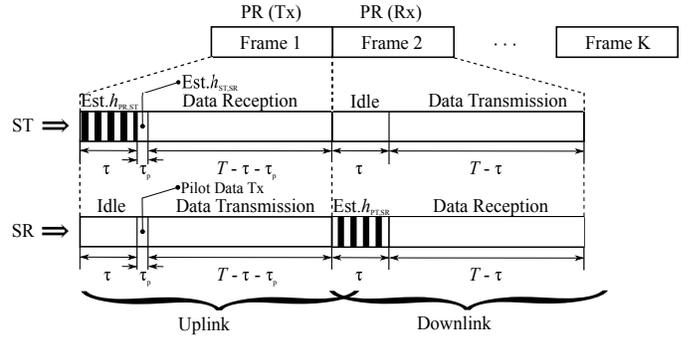}
\vspace{\figmarc}
\caption{\tcg{Frame structure of the US illustrating the time allocation for channel estimation and data transmission from the perspective of a ST and a SR. In this regard, corresponding to the uplink and the downlink, the primary interference and secondary interference channel estimation occur at the ST and the SR, respectively. PR (Tx)/PR (Rx) represents the transmission/reception of the primary signal from the PR/PT to the PT/PR.}} 
\label{fig:fs}
\vspace{\figmarb}
\end{figure}

We propose to employ a slotted medium access for the US, where the time axis is segmented into frames. As depicted in \figurename~\ref{fig:fs}, the frame duration $T$ is chosen in such a way that the frames are aligned to the primary users' transmissions\tcg{, i.e., the uplink and the downlink transmissions for the primary and secondary systems occur simultaneously. In this regard, a perfect frame synchronization is assumed between the two systems.} In order to incorporate channel estimation, we further propose to employ a periodic channel estimation\footnote{This frame structure is similar to the periodic sensing followed by the interweave systems \cite{Liang08}.}, according to which the US uses time intervals \tcg{($\tau$ and $\tau\sub{p}$ in the uplink, and $\tau$ in the downlink)} to perform channel estimation followed by data transmission \tcg{($T- \tau - \tau\sub{p}$ in the uplink and $T - \tau$ in the downlink)}, see \figurename~\ref{fig:fs}. In order to consider variations due to channel fading, we assume that the interacting channels remain constant over at least two frame durations ($2T$). Based on this assumption, every alternating transmission frame observes a different received power, consider \figurename~\ref{fig:fs}. Since the channel knowledge is essential to employ the power control so that the primary users are sufficiently protected from the uncertain interference induced due to the imperfect channel knowledge, it is reasonable to carry out estimation for $\tau$ time interval followed by data transmission with controlled power in the remaining time for each frame. 

In accordance with the half duplexing modes, the ST and the SR implement the received power-based estimation to acquire the knowledge of the primary and the secondary interference channel over consecutive frames, as illustrated in \figurename~\ref{fig:fs}. \tcg{For the case where the primary system follows full-duplex FDD,} the proposed frame structure can be adapted such that the primary and the secondary interference channel estimation occurs in a single frame. Besides this, the access channel estimation is performed by listening to the pilot symbols transmitted by the SR, classified as the pilot-based channel estimation. 
At first, we consider the proposed frame structure for a deterministic channel, i.e., the performance is analyzed for a certain channel gain (path-loss channel), without taking into account the effect of channel fading. \tcg{We then extend} the performance analysis for the proposed framework by considering channel fading. 


\subsection{Signal Model}
\tc{In the uplink,} during the estimation phase, the discrete and complex signal received from the PR at the ST is given by
\tc{\begin{equation}
\yrcvd[n] = \gpo \cdot \sqrt{\ptran} \cdot \xtran[n] + \nas[n],
\label{eq:sys_mod_st}
\end{equation}}
where $\xtran[n]$ corresponds to a discrete and complex sample transmitted by the PR with unit power, $\ptran$ is the transmit power at the PR (known at the ST), $\pgpo$ represents the power gain for the primary interference channel and $\nas[n]$ is circularly symmetric Additive White Gaussian Noise (AWGN) at the ST with 
$\mathcal{CN}(0, \nps)$. 

\tc{In the downlink,} during data transmission phase, the \tcg{interference (received from the ST) plus noise} signal at the PR is given by
\tc{\begin{equation}
\yp[n] = \gpo  \cdot \sqrt{\preg} \cdot \xreg[n] + \nap[n],
\label{eq:sys_mod_pr}
\end{equation}}
and on the other side, the received signal at the SR follows 
\tc{\begin{align}
\ys[n] =& \gs \cdot \sqrt{\preg} \cdot \xreg[n] +  \nonumber \\ & \gpt \cdot \sqrt{\ptranpt} \cdot \xtranpt[n] + \nasr[n],
\label{eq:sys_mod_sr}
\end{align}
}where $\xreg[n]$ corresponds to a discrete and complex sample transmitted by the ST with unit power, $\preg$ is the controlled transmit power and $\xtranpt[n]$ is the transmit signal from the PT with \tcg{transmit power $\ptranpt$}\footnote{\tcg{
In reference to the proposed framework, the knowledge of the PT's transmit power is not necessary at the secondary system. Hence, its ignorance at the SR does not affect the analysis concerning the secondary interference channel. 
With no loss of generality, for our analysis, the PT and the PR are alloted the same transmit power.}}. Further, $\pgs$ and $\pgpt$ represent the power gain for the access channel and the secondary interference channel, respectively. $\nap[n]$ and $\nasr[n]$ are AWGN at the PR and at the SR, respectively, with $\mathcal{CN}(0, \nps)$\footnote{\tc{In practice, the noise power at the ST, the SR and the PR has different values. The fact is, only the signal to noise ratio received at the ST, and SR and the PR, respectively, are affected due to these noise powers. Since these signal to noise ratios are already included in the performance analysis, \tcg{the use of different notations to these noise powers in the expressions is avoided.}}}. 


\subsection{Problem Description} \label{ssec:pd}
According to the existing investigations (also referred as ideal model), an ST of an US is required to control its transmit power in such a way that the interference received ($\pp$) at the PR is below IT ($\ite$) \cite{Xing07}
\begin{equation}
\pp = \pgpo \preg \le \ite.
\label{eq:IT_id}
\end{equation}

\tcg{After determining the controlled power} at the ST using (\ref{eq:IT_id}), the data rate at the SR over the access channel is defined as
\begin{equation}
\ca = \log_2 \left(1 + \frac{\pgs \preg}{ \pgpt \ptranpt + \nps} \right). 
\label{eq:Thr_id}
\end{equation}
From the deployment perspective, the ideal model depicted in (\ref{eq:IT_id}) and (\ref{eq:Thr_id}) has following issues:
\begin{itemize}
\item Without the knowledge of the primary interference channel $\gpo$, it is impossible to employ the power control at the ST, which is based on (\ref{eq:IT_id}). 
\item Furthermore, along with $\preg$, the knowledge of the access channel $\gs$ and the secondary interference channel $\gpt$ is required to determine $\ca$, according to (\ref{eq:Thr_id}).
\end{itemize}
The ideal model considers the perfect knowledge of the aforementioned channels at the ST, which is not available in practice. In this regard, it is necessary to incorporate channel estimation in the system model. The imperfect channel knowledge, however, translates to the variations in the performance parameters, $\pp$ and $\ca$. \tcg{Particularly, a variation in $\pp$ due to uncertain interference that exceeds $\ite$ causes the violation of the outage constraint illustrated in (\ref{eq:IT_id}).} Unless captured, this uncertain interference may seriously degrade the performance of the US. Since the ideal model assumes the perfect knowledge of the involved channels, it is incapable of depicting the degradation in the performance due to the time allocation for the channel estimation and the imperfect knowledge of the channels. 

\subsection{Proposed Approach} 
In order to facilitate channel estimation for the US, it is essential to take the aforementioned issues into account. To accomplish this, the following strategy is proposed in the paper.
\begin{itemize}
\item At first, we consider the estimation of the involved channels. In this regard, we propose to employ a received power-based estimation for the interference channels and a pilot-based estimation for the access channel. 
\item To capture the effect of the imperfect channel knowledge, we characterize the variations in the estimated parameters (namely, received power for the interference channels and power gain for the access channels) in terms of their cdfs.
\item \tcg{The aforementioned variations are translated to the performance parameters, which include the uncertain interference and the secondary throughput. We further characterize these variation in the performance parameters in terms of their cdfs. More specifically, using the characterization of the uncertain interference, we propose a novel power control mechanism that regulates the uncertain interference at the PR.} 
\item Finally, using the derived expressions, we analyze a relationship between the estimation time and the expected secondary throughput for the US. We extend the proposed framework (also referred as estimation model) to analyze the impact of channel fading on the performance of the system. 
\end{itemize}
Since the channel estimation in the context of CR systems involves different systems, suitable channel estimation techniques should be selected such that the following requirements: (i) low complexity and (ii) versatility towards unknown primary user signals, essential from the deployment perspective, are respected. Similar problem for the interweave CR systems has been deeply investigated in \cite{Kaushik16}, where the authors propose to employ a received power-based estimation for the channels between the primary and the secondary systems and a pilot-based estimation for the access channel. Following a similar strategy, we propose to employ received power-based and pilot-based estimation techniques in the underlay CR systems. It is also worth stating that, since the signal model in \cite{Kaushik16} (orthogonal frequency division multiplexing transmission) differs from the one (constant power transmission) studied in this paper, we derive new mathematical expressions for the performance parameters. 
In the following paragraphs, we consider the estimation of the power gains of the primary interference channel $\epgpo$, the access channel $\epgs$ and the secondary interference channel $\epgpt$. 
\subsubsection{Estimation of primary interference channel}
\tc{Considering}
\begin{equation}
\prcvd = \pgpo \ptran + \nps \label{eq:prcvd}, 
\end{equation}
\tc{and the knowledge of PR's transmit power $\ptran$, the ST employs received power-based estimation to obtain the knowledge of $\pgpo$. To accomplish this, in reference to (\ref{eq:sys_mod_st}), the ST listens to the transmissions from the PR and acquires the knowledge of $\pgpo$ indirectly by estimating the power received in the uplink as $\eprcvd = \s{\tau \fsam}{|\yrcvd[n]|^2}$, where $\fsam$ being the sampling frequency and $\tau$ represents the estimation time interval. $\fsam$ and $\tau$ are such that the number of samples $\tau \fsam$ is an integer. The estimated received power $\eprcvd$ is utilized to determine the controlled power $\preg$ at which the data transmission over the downlink is carried out, consider \figurename~\ref{fig:fs}. In accordance to the received power-based estimation for the primary interference channel in (\ref{eq:prcvd}), it is noticed that the knowledge of $\ptran$ at the ST is essential for the characterization of the controlled power (considered later in Lemma \ref{lm:lm4}).} This knowledge can be retrieved from the specification of different wireless standards such as GSM, EDGE and LTE, etc. \cite{Sharma14}. It is well-known that certain standards follow adaptive modulation and coding, which can consequently change $\ptran$. Under this situation, the ST can employ more complex techniques such as pilot assisted techniques in order to determine $\ptran$ for the given frame.

\tc{For a certain value of $\pgpo$, the received power at the ST} estimated using $\tau \fsam$ samples follows a non-central chi-squared distribution $\fprcvd \sim \ncchi2(\lpo, \tau \fsam)$ with non-centrality parameter $\lpo = \tau \fsam \pgpo \ptran /\nps = \tau \fsam \gamma$ \cite{Kay}, where $\gamma$ is defined as the ratio of the received signal power (from the PR) to noise at the ST and $\tau \fsam$ corresponds to the degrees of freedom. For analytical tractability, we consider the following approximation. 
\begin{approxi} \label{ap:ap1}
\normalfont
For all degrees of freedom, the $\ncchi2$ distribution can be approximated by a Gamma distribution \cite{abramo}. The parameters of the Gamma distribution are obtained by matching the first two central moments to those of $\ncchi2$.
\end{approxi}
\begin{lemma} \label{lm:lm1}
\normalfont
The cdf of $\eprcvd$ is characterized as 
\begin{align}
\fprcvd(x) \approx 1 - \Gamma&\left(\apo, \frac{x}{\bpo}\right) \label{eq:fprcvd}, \\ 
\text{where  } \apo = \frac{\tau \fsam (1 + \gamma)^2}{2 + 4 \gamma} &\text{ and } \bpo = \frac{\nps (2 + 4 \gamma)}{\tau \fsam (1 + \gamma)},  \label{eq:para_po} 
\end{align} 
and $\Gamma(\cdot, \cdot)$ represents the regularized upper-incomplete Gamma function \cite{abramo}. 
\end{lemma}
\begin{IEEEproof}
Applying Approximation \ref{ap:ap1} to $\ncchi2(\lpo, \tau \fsam)$ yields (\ref{eq:fprcvd}). 
\end{IEEEproof}

\subsubsection{Estimation of access channel}
In the uplink, the discrete and complex pilot signal \tc{transmitted by the SR} undergoes matched filtering and demodulation at the ST, consider \figurename~\ref{fig:fs}, hence, we employ a pilot-based estimation at the ST to acquire the knowledge of the access channel. According to \cite{Gifford08}, the maximum-likelihood estimate with $\Ks$ pilot symbols is given by 
\begin{align}
\egs = \gs + \frac{\sum^{\Ks}_{n = 1} \nas[n]}{\Ks},
\label{eq:pilot_MLE}
\end{align}
where $\frac{\sum^{\Ks}_{n} \nas[n]}{\Ks}$ represents the estimation error. With no loss of generality, the pilot symbols are considered to be $+1$.
As a result, the estimate $\egs$ is unbiased, efficient, i.e., achieves the Cram\'er-Rao bound with equality, with asymptotic variance $\e{}{|\gs -\egs|^2} = \frac{\nps}{\Ks}$ \cite{Gifford08}. Hence, $\egs$ conditioned on $\gs$ follows a circularly symmetric Gaussian distribution, given by
\begin{align}
\egs|\gs \sim \mathcal{C} \mathcal{N}\left( \gs,\frac{\nps}{\Ks} \right).
\label{eq:ehs} 
\end{align}
Consequently, \tc{for a certain value of $\pgs$,} the estimated power gain $|\egs|^2$ follows a non-central chi-squared $\ncchi2(\ls, 2)$ distribution with 2 degrees of freedom and non-centrality parameter $\ls = \frac{\Ks |\gs|^2}{\nps}$. 
\begin{lemma} \label{lm:lm2}
\normalfont
The cdf of $\epgs$ is characterized as 
\begin{align}
\fgs(x) \approx 1 - \Gamma&\left(\as, \frac{x}{\bs}\right) \label{eq:fehs},\\ 
\text{where  } \as = \frac{(2 + \ls)^2}{4 + 4 \ls} &\text{ and } \bs = \frac{\nps (4 + 4 \ls)}{(2 + \ls)}. \label{eq:para_s} 
\end{align} 
\end{lemma}
\begin{IEEEproof}
Applying Approximation \ref{ap:ap1} to $\ncchi2(\ls, 2)$ yields (\ref{eq:fehs}). 
\end{IEEEproof}

\subsubsection{Estimation of secondary interference channel}
\tcg{In the downlink, the SR estimates the interference (power) received from the PT, consider (\ref{eq:sys_mod_sr})}. The power estimated over the signal \tcg{$\gpt \cdot \sqrt{\ptranpt} \cdot \xtranpt[n] + \nasr[n]$} corresponds to the interference plus noise power ($\prcvdsr = \pgpt \ptran + \nps$, where $\prcvdsr$ represents the true value, consider (\ref{eq:Thr_id})). The estimated received power at the SR is determined as \tcg{$\eprcvdsr = \s{\tau \fsam}{|\gpt \cdot \sqrt{\ptranpt} \cdot \xtranpt[n] + \nasr[n]|^2}$.} To characterize the secondary throughput, $\eprcvdsr$ is made available to the ST over a low rate feedback channel. Similar to $\eprcvd$, \tc{for a certain value of $\pgpt$}, $\eprcvdsr$ follows a non-central chi-squared distribution \tc{$\ncchi2(\lpt, \tau \fsam)$}, with non-centrality parameter $\lpt = \tau \fsam \pgpt \ptranpt / \nps$.
\begin{lemma} \label{lm:lm3}
\normalfont
The cdf of $\eprcvdsr$ is characterized as 
\begin{align}
\fprcvdsr(x) \approx 1 - \Gamma&\left(\apt, \frac{x}{\bpt}\right) \label{eq:fprcvdsr}, \\ 
\text{where  } \apt = \frac{(\tau \fsam + \lpt)^2}{2 \tau \fsam + 4 \lpt} &\text{ and } \bpt = \frac{\nps (2 \tau \fsam + 4 \lpt)}{(\tau \fsam + \lpt)}.  \label{eq:para_pt} 
\end{align} 
\end{lemma}
\begin{IEEEproof}
Applying Approximation \ref{ap:ap1} to $\ncchi2(\lpt, \tau \fsam)$ yields (\ref{eq:fprcvdsr}). 
\end{IEEEproof}
It is important to note that, in this paper, we are dealing with a single PT and a single PR. However, in practice, it is possible that the ST and the SR accumulate significant interference (defined as aggregate interference) from other PRs and PTs (co-channel interference due to frequency reuse) in the network\cite{Elsawy13_cmag,Kaushik14_P} over the primary interference channel and the secondary interference channel, respectively. For the secondary interference channel, the only difference is that the SR now estimates the aggregate interference. Due to this, the expression of $\eprcvdsr$ in the secondary throughput remains unchanged. On the other side, by estimating the aggregate interference on the primary interference channel, the ST overestimates $\eprcvd$ and exercises a greater power control. Even for such a case, the outage constraint on the primary interference channel to the desired PR is satisfied, however, reduces the secondary throughput.  

\section{Theoretical Analysis} \label{sec:th_ana}

\subsection{Deterministic Channel} \label{ssec:stpa}
In this section, we investigate the performance of the US \tcg{for a specific frame}. In this sense, the involved channels $\gpo$, $\gpt$ and $\gs$ are deterministic \tc{(not random)}.   
First, we employ an outage probability constraint\footnote{The outage constraint is commonly used parameter for designing communication system that ensures the outage occurs no more than a certain percentage of time.} $\opc$ on the interference to capture the variations in the $\pp$ incurred due to channel estimation, defined as 
\begin{align}
\tc{\p\left( \pp = \epgpo \preg \ge \ite \right) \le \opc.} 
\intertext{Substituting $\epgpo$ from (\ref{eq:prcvd}) yields}
\p\left( \left( \frac{\eprcvd - \nps}{\ptran}\right) \preg \ge \ite \right) \le \opc. \label{eq:opc} \\[-1em] \nonumber 
\end{align}
Besides the outage constraint, $\preg$ is limited by a predefined transmit power $\pc$. To capture this aspect, the transmit power constraint at the ST is defined as
\begin{align}
\preg \le \pc. \label{eq:pc} 
\end{align} 
We consider that the same power is allocated to all the symbols transmitted within a frame by the ST. In this regard, the transmit power constraint on symbol basis and frame basis is equivalent. As a consequence, the constraint depicted in (\ref{eq:pc}) \tcg{is applicable to both the cases.} Based on the constraints in (\ref{eq:opc}) and (\ref{eq:pc}), we subsequently determine the expression of the controlled power for the proposed framework.
\begin{lemma} \label{lm:lm4}
\normalfont 
Subject to the outage constraint on the uncertain interference and the transmit power constraint at the ST, the controlled power at the ST is given by
\begin{align}
\preg &= 
\begin{cases} 
\frac{\ite \ptran}{ \left(\bpo \Gamma^{-1}(\opc, \apo) - \nps  \right)}, & \mbox{for } \preg < \pc \\
\pc, & \mbox{for } \preg \ge \pc
\end{cases},
\label{eq:preg} 
\end{align}
where $\apo$ and $\bpo$ are defined in (\ref{eq:para_po}) and $\Gamma^{-1}(\cdot, \cdot)$ is the inverse function of regularized upper-incomplete Gamma function \cite{abramo}.
\end{lemma} 
\begin{IEEEproof}
Substituting the cdf $\fprcvd(x)$, defined in (\ref{eq:fprcvd}), in (\ref{eq:opc}) and combining with (\ref{eq:pc}) yields (\ref{eq:preg}).
\end{IEEEproof}
\tc{Clearly, the performance of the US improves over the access channel (in terms of the secondary throughput) with $\preg$, but $\preg$ increases for the values of $\pgpo$, which correspond to the lower values of $\gamma$\footnote{\tc{Signal to noise ratio is mostly used as a design parameter for characterizing the performance of a wireless system.}}. \tcg{However, in practice, the wireless systems} are limited by the transmit power $\pc$, which bounds the performance of the US. In order to understand the effect of the power limitation on the US, we characterize a performance bound in terms of the estimation time.}
\begin{coro} \label{cor:cor1}
\normalfont
Subject to the outage constraint on the uncertain interference and the transmit power constraint at the ST, the performance bound ($\gamma^*$) of the US is defined as 
\begin{align}
\Gamma\left(\frac{\tau \fsam (1 + \gamma^*)^2}{2 + 4 \gamma^*}, \frac{\tau \fsam (1  + \gamma^*)}{\nps (2 + 4 \gamma^*)} \left( \frac{\ite \ptran}{\pc} + \nps  \right)  \right) = \opc. \label{eq:opreg}  
\end{align}
\end{coro}
\begin{IEEEproof}
Substituting $\preg$ with $\pc$ in (\ref{eq:opc}) and reformulating gives 
\begin{align}
\p\left( \eprcvd \ge \frac{\ite \ptran}{\pc} + \nps \right) &\le \opc. \label{eq:opreg_a} 
\intertext{Using (\ref{eq:fprcvd}) in Lemma \ref{lm:lm1} gives}
\Gamma\left(\frac{\tau \fsam (1 + \gamma)^2}{2 + 4 \gamma}, \frac{\tau \fsam (1  + \gamma)}{\nps (2 + 4 \gamma)} \left( \frac{\ite \ptran}{\pc} + \nps  \right)  \right) &\le \opc. \label{eq:opreg_b} 
\end{align}
Substituting $\gamma$ with $\gamma^*$ and replacing the expression in (\ref{eq:opreg_b}) with equality yields (\ref{eq:opreg}). 
\end{IEEEproof}
\begin{figure}[!t]
\vspace{\figmara}
%
%
%
\psfrag{s03}[b][b]{\fontsize{8.5}{12.75}\fontseries{m}\mathversion{normal}\fontshape{n}\selectfont \color[rgb]{0,0,0}\setlength{\tabcolsep}{0pt}\begin{tabular}{c}$\tau$ [ms]\end{tabular}}%
\psfrag{s04}[t][t]{\fontsize{8.5}{12.75}\fontseries{m}\mathversion{normal}\fontshape{n}\selectfont \color[rgb]{0,0,0}\setlength{\tabcolsep}{0pt}\begin{tabular}{c}$\gamma$ [dB]\end{tabular}}%
%
\fontsize{8.5}{12.75}\fontseries{m}\mathversion{normal}%
\fontshape{n}\selectfont%
%
\psfrag{x01}[t][t]{-20}%
\psfrag{x02}[t][t]{-15}%
\psfrag{x03}[t][t]{-10}%
\psfrag{x04}[t][t]{-5}%
%
\psfrag{v01}[r][r]{0}%
\psfrag{v02}[r][r]{2}%
\psfrag{v03}[r][r]{4}%
\psfrag{v04}[r][r]{6}%
\psfrag{v05}[r][r]{8}%
\psfrag{v06}[r][r]{10}%
\psfrag{v07}[r][r]{12}%
\psfrag{v08}[r][r]{14}%
\psfrag{v09}[r][r]{16}%
\psfrag{v10}[r][r]{18}%
\psfrag{v11}[r][r]{20}%
%
%

\centering
\begin{tikzpicture}[scale=1]
\node[anchor=south west,inner sep=0] (image) at (0,0)
{
\includegraphics[width= \figscale]{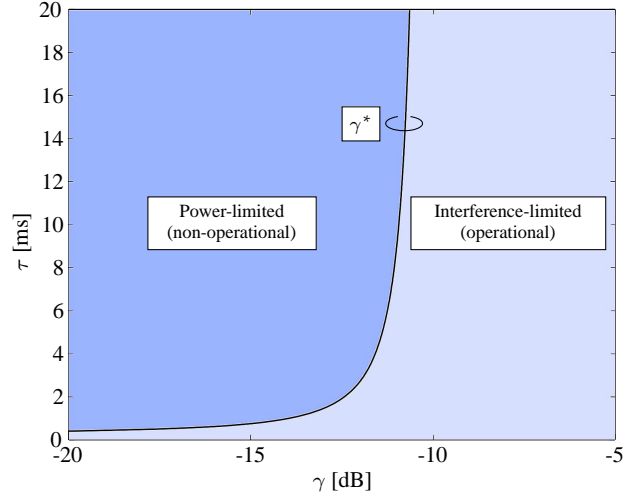}
};
\begin{scope}[x={(image.south east)},y={(image.north west)}]


\draw (0.62,0.75) arc(-250:70:0.03 and 0.015);
\node[draw,fill=gray!0,font=\scriptsize] (text1) at (0.56,0.735) {$\gamma^*$}; 
\node[draw,fill=gray!0,font=\scriptsize, text width = 68, align = center] at (0.8,0.53) {Interference-limited (operational)};
\node[draw,fill=gray!0,font=\scriptsize, text width = 58, align = center] at (0.35,0.53) {Power-limited (non-operational)};

\end{scope}
\end{tikzpicture}
\vspace{\figmard}
\caption{An illustration of the performance bound ($\gamma^*$) for the US depicted in terms of estimation time ($\tau$), where $\gamma$ represents the ratio of the received power (from the PR) to noise at the ST. It further classifies the operation of the US as the interference-limited and the power-limited regimes.}
\label{fig:or}
\vspace{\figmarb}
\end{figure}
\begin{remark} \label{rm:rm1}
\normalfont
\figurename~\ref{fig:or} analyzes the variations of $\gamma^*$ with $\tau$. Using the expression $\gamma^*$ obtained in Corollary \ref{cor:cor1}, we classify the operation of the US into the following regimes: (i) interference-limited regime and (ii) power-limited regime. \tc{Inside the interference-limited regime $\gamma > \gamma^*$, due to good quality of the channel ST-PR (unfavorable to the US), the system is limited due to the exceeding level of the uncertain interference, which can be regulated effectively by employing power control at the US to satisfy the given outage constraint ($\opc$). At $\gamma = \gamma^*$, the ST operates at the maximum allowable power $\preg = \pc$ while respecting the tolerance limits defined for the uncertain interference. From a different perspective, the situation $\gamma = \gamma^*$ also represents those USs that are unable to carry out power control. With regard to the outage constraint and the lack of the power control, for a given choice of $\gamma^*$, such systems can operate only at a specific value of $\tau$.
}

\tc{
On the other side, the region $\gamma < \gamma^*$, which depicts a weak link quality between the ST and the PR, is beneficial to the secondary user. However, due to the transmit power constraint, the USs \tcg{can operate at or below $\pc$}. As a result, these favorable conditions do not translate to any performance gain.} Therefore, this regime is characterized as a power-limited regime. \tc{Besides, it is interesting to observe that for small values of the estimation time, $\gamma^* \rightarrow -\infty$, which signifies that low $\tau$ increases the uncertainty in the interference. In order to regulate the level of this uncertainty, US has to be proactive in terms of the power control to be able to satisfy the outage constraint. It is also observed that as $\tau \rightarrow \infty$, $\gamma^*$ converges asymptotically to a certain value. \tcg{This signifies the fact that, beyond a certain value, the time resources allocated for the channel estimation} do not account to any significant improvement in the terms of the uncertain interference or indirectly in terms of the controlled power. As a result, the performance of the US in the form of controlled power gets saturated, thus, limiting the performance of the US in terms of the secondary throughput.}  
\end{remark}

Next, we capture the variations in the secondary throughput in terms of its expected value. To accomplish this, the cdf of the estimated data rate\footnote{\tc{Please note, we have introduced the following terms the data rate $\ca$ and the throughput $\rs$ to make a clear distinction between the instantaneous data rate and its average value over the frame duration.}}, given by  
\begin{align}
\eca  = \log_2 \left(1 + \frac{\epgs \preg}{\eprcvdsr} \right) \label{eq:eca}
\end{align}
is evaluated over the access channel at the ST. It is worth noticing the fact that unlike $\ca$ defined in (\ref{eq:Thr_id}), $\eca$ entails the random behavior due to the estimation of $\epgs$ and $\eprcvdsr$.
\begin{lemma} \label{lm:lm5}
\normalfont 
The cdf of data rate $\eca$ is given by
\begin{align}
\fc(x) &= \int\limits_{0}^{x} \dc(t) dt, \label{eq:dis_C} 
\end{align}
where the pdf is given by 
\begin{align}
\dc(x) =& 2^x \ln 2 \frac{(2^x - 1)^{\as - 1} \Gamma(\as + \apt)}{\Gamma(\as) \Gamma(\apt) (\bs \preg) ^{\as} \bpt^{\apt}} \times \nonumber \\  & \left(\frac{1}{\bpt} + \frac{2^x - 1}{\bs \preg}\right). \label{eq:den_C}
\end{align}
\end{lemma}
\begin{IEEEproof}
See Appendix \ref{ap:one}.
\end{IEEEproof}

\captionsetup[subfigure]{position=top}
\begin{figure*}[!t]
\vspace{\figmara}
\centering
\subfloat[]{
%
%
%
\psfrag{s05}[t][t]{\fontsize{8}{12}\fontseries{m}\mathversion{normal}\fontshape{n}\selectfont \color[rgb]{0,0,0}\setlength{\tabcolsep}{0pt}\begin{tabular}{c}$\eca$ [bits/sec/Hz]\end{tabular}}%
\psfrag{s06}[b][b]{\fontsize{8}{12}\fontseries{m}\mathversion{normal}\fontshape{n}\selectfont \color[rgb]{0,0,0}\setlength{\tabcolsep}{0pt}\begin{tabular}{c}CDF\end{tabular}}%
\psfrag{s10}[][]{\fontsize{10}{15}\fontseries{m}\mathversion{normal}\fontshape{n}\selectfont \color[rgb]{0,0,0}\setlength{\tabcolsep}{0pt}\begin{tabular}{c} \end{tabular}}%
\psfrag{s11}[][]{\fontsize{10}{15}\fontseries{m}\mathversion{normal}\fontshape{n}\selectfont \color[rgb]{0,0,0}\setlength{\tabcolsep}{0pt}\begin{tabular}{c} \end{tabular}}%
\psfrag{s12}[l][l]{\fontsize{8}{12}\fontseries{m}\mathversion{normal}\fontshape{n}\selectfont \color[rgb]{0,0,0}$\frac{\pgpt\ptran}{\nps}=-10$dB}%
\psfrag{s13}[l][l]{\fontsize{8}{12}\fontseries{m}\mathversion{normal}\fontshape{n}\selectfont \color[rgb]{0,0,0}$\frac{\pgpt\ptran}{\nps}=10$dB}%
\psfrag{s14}[l][l]{\fontsize{8}{12}\fontseries{m}\mathversion{normal}\fontshape{n}\selectfont \color[rgb]{0,0,0}$\frac{\pgpt\ptran}{\nps}=0$dB}%
\psfrag{s15}[l][l]{\fontsize{8}{12}\fontseries{m}\mathversion{normal}\fontshape{n}\selectfont \color[rgb]{0,0,0}$\frac{\pgpt\ptran}{\nps}=-10$dB}%
%
\fontsize{8}{12}\fontseries{m}\mathversion{normal}%
\fontshape{n}\selectfont%
%
\psfrag{x01}[t][t]{3}%
\psfrag{x02}[t][t]{3.5}%
\psfrag{x03}[t][t]{4}%
\psfrag{x04}[t][t]{4.5}%
\psfrag{x05}[t][t]{5}%
\psfrag{x06}[t][t]{5.5}%
\psfrag{x07}[t][t]{6}%
\psfrag{x08}[t][t]{6.5}%
%
\psfrag{v01}[r][r]{0}%
\psfrag{v02}[r][r]{0.1}%
\psfrag{v03}[r][r]{0.2}%
\psfrag{v04}[r][r]{0.3}%
\psfrag{v05}[r][r]{0.4}%
\psfrag{v06}[r][r]{0.5}%
\psfrag{v07}[r][r]{0.6}%
\psfrag{v08}[r][r]{0.7}%
\psfrag{v09}[r][r]{0.8}%
\psfrag{v10}[r][r]{0.9}%
%
%

\begin{tikzpicture}[scale=1]
\node[anchor=south west,inner sep=0] (image) at (0,0)
{
        \includegraphics[width = \figscale]{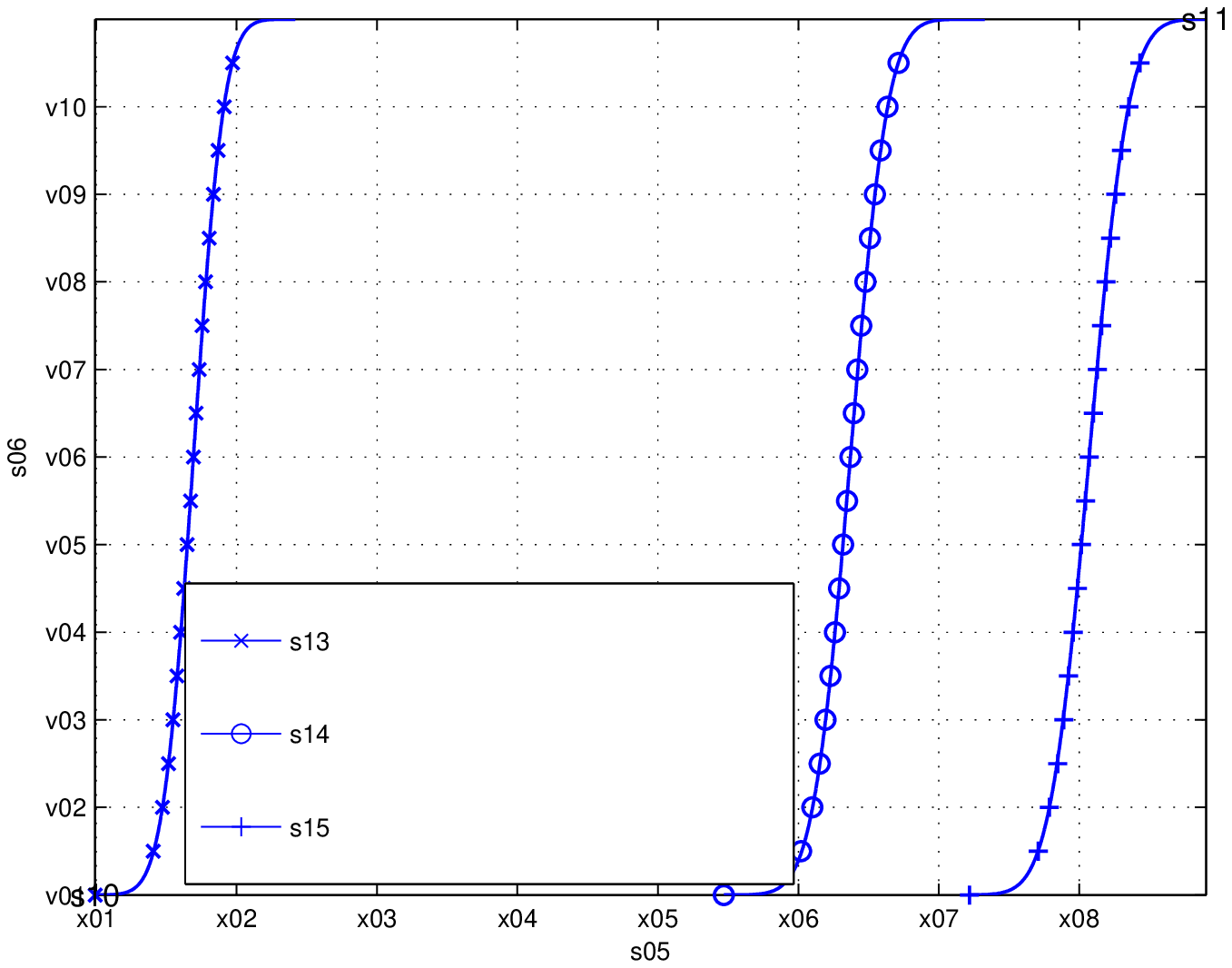} 
};
\begin{scope}[x={(image.south east)},y={(image.north west)}]


\end{scope}
\end{tikzpicture}

\label{fig:CDF_C1_s}}
\hfil
\subfloat[]{
%
%
%
\psfrag{s05}[t][t]{\fontsize{8}{12}\fontseries{m}\mathversion{normal}\fontshape{n}\selectfont \color[rgb]{0,0,0}\setlength{\tabcolsep}{0pt}\begin{tabular}{c}$\eca$ [bits/sec/Hz]\end{tabular}}%
\psfrag{s06}[b][b]{\fontsize{8}{12}\fontseries{m}\mathversion{normal}\fontshape{n}\selectfont \color[rgb]{0,0,0}\setlength{\tabcolsep}{0pt}\begin{tabular}{c}CDF\end{tabular}}%
\psfrag{s10}[][]{\fontsize{10}{15}\fontseries{m}\mathversion{normal}\fontshape{n}\selectfont \color[rgb]{0,0,0}\setlength{\tabcolsep}{0pt}\begin{tabular}{c} \end{tabular}}%
\psfrag{s11}[][]{\fontsize{10}{15}\fontseries{m}\mathversion{normal}\fontshape{n}\selectfont \color[rgb]{0,0,0}\setlength{\tabcolsep}{0pt}\begin{tabular}{c} \end{tabular}}%
\psfrag{s12}[l][l]{\fontsize{8}{12}\fontseries{m}\mathversion{normal}\fontshape{n}\selectfont \color[rgb]{0,0,0}$\tau = 10$ ms}%
\psfrag{s13}[l][l]{\fontsize{8}{12}\fontseries{m}\mathversion{normal}\fontshape{n}\selectfont \color[rgb]{0,0,0}$\tau = 0.1$ ms}%
\psfrag{s14}[l][l]{\fontsize{8}{12}\fontseries{m}\mathversion{normal}\fontshape{n}\selectfont \color[rgb]{0,0,0}$\tau = 1$ ms}%
\psfrag{s15}[l][l]{\fontsize{8}{12}\fontseries{m}\mathversion{normal}\fontshape{n}\selectfont \color[rgb]{0,0,0}$\tau = 10$ ms}%
%
\fontsize{8}{12}\fontseries{m}\mathversion{normal}%
\fontshape{n}\selectfont%
%
\psfrag{x01}[t][t]{6}%
\psfrag{x02}[t][t]{6.2}%
\psfrag{x03}[t][t]{6.4}%
\psfrag{x04}[t][t]{6.6}%
\psfrag{x05}[t][t]{6.8}%
\psfrag{x06}[t][t]{7}%
\psfrag{x07}[t][t]{7.2}%
\psfrag{x08}[t][t]{7.4}%
%
\psfrag{v01}[r][r]{0}%
\psfrag{v02}[r][r]{0.1}%
\psfrag{v03}[r][r]{0.2}%
\psfrag{v04}[r][r]{0.3}%
\psfrag{v05}[r][r]{0.4}%
\psfrag{v06}[r][r]{0.5}%
\psfrag{v07}[r][r]{0.6}%
\psfrag{v08}[r][r]{0.7}%
\psfrag{v09}[r][r]{0.8}%
\psfrag{v10}[r][r]{0.9}%
%
%
 
\begin{tikzpicture}[scale=1]
\node[anchor=south west,inner sep=0] (image) at (0,0)
{
        \includegraphics[width = \figscale]{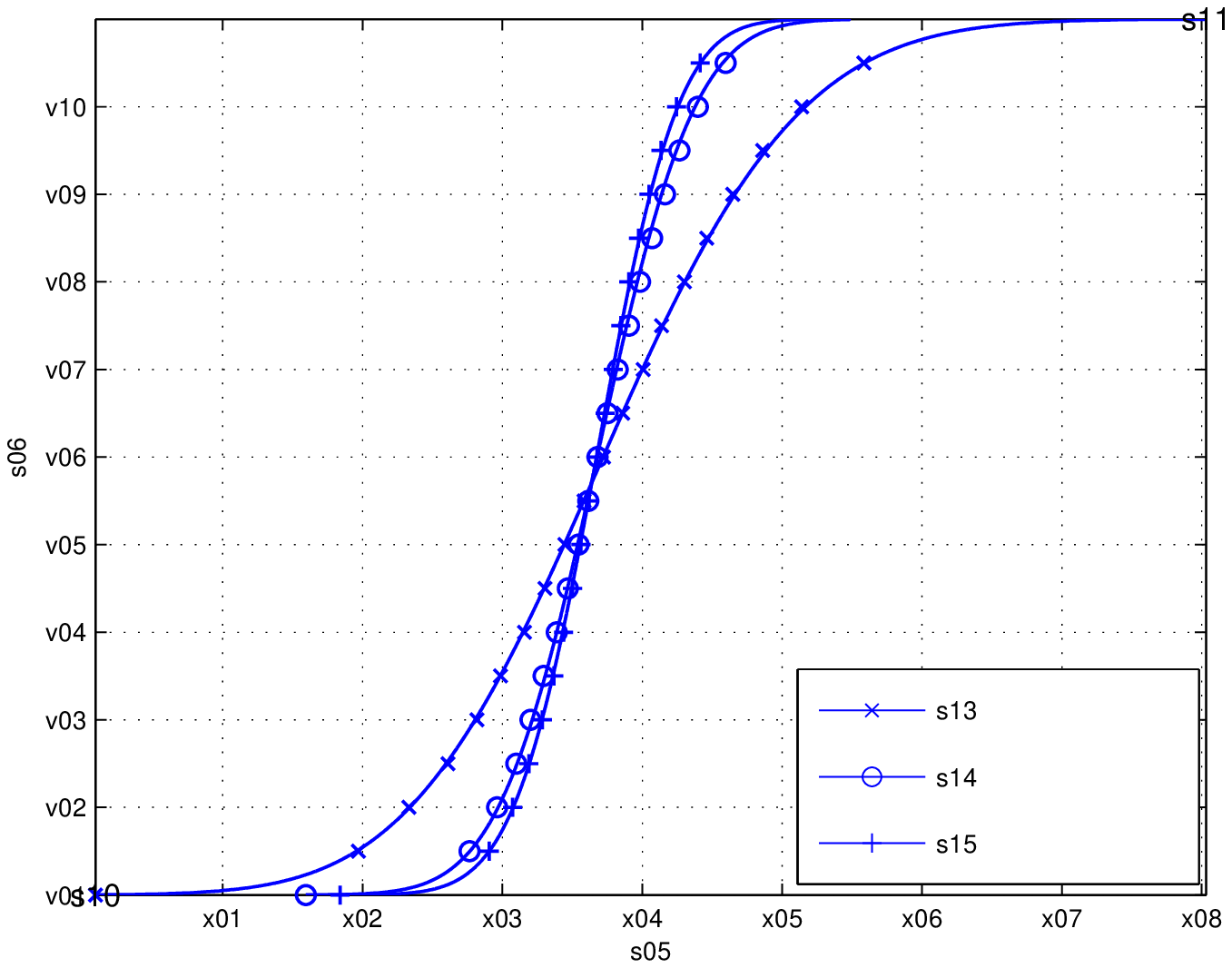} 
};
\begin{scope}[x={(image.south east)},y={(image.north west)}]


\end{scope}
\end{tikzpicture}
\label{fig:CDF_C1_p2}}
\vspace{\subfigmar}
\caption{CDF of $\eca$ for different received \tcg{interference} to noise ratio \tcg{$\frac{\pgpt \ptranpt}{\nps}$} over the secondary interference channel and estimation time $\tau$. (a) \tcg{$\frac{\pgpt \ptranpt}{\nps} \in \{-10, 0, 10\} \SI{}{dB}$} and $\tau = \SI{1}{ms}$, (b) $\tau = \{0.1, 1, 10\} \SI{}{ms}$ and \tcg{$\frac{\pgpt \ptran}{\nps} = \SI{0}{dB}$}, for the following values of the system parameters, $\gamma = \SI{10}{dB}$, $\preg = \SI{0}{dBm}$. \tc{The solid line represents the values computed using the analytical expressions while the markers represent the values obtained from the simulations.}}
\label{fig:CDF_eca}
\vspace{\figmarb}
\end{figure*}
\tc{In consideration to the Approximation \ref{ap:ap1}, which is applied to obtain the cdfs' of $\eprcvd$, $\epgs$ and $\eprcvdsr$ in Lemma \ref{lm:lm5}, the theoretical expression of the cdf depicted in (\ref{eq:dis_C}) is validated by means of simulations in \figurename~\ref{fig:CDF_eca} with different choices of the system parameters, which include $\gamma = \SI{10}{dB}$, $\preg = \SI{0}{dBm}$, \tcg{$\frac{\pgpt \ptranpt}{\nps} \in \{-10, 0, 10\} \SI{}{dB}$} and $\tau = \{0.1, 1, 10\} \SI{}{ms}$.} 

Besides the outage constraint on the \tcg{uncertain interference}, the expected secondary throughput \tcg{(averaged over $2T$ frame durations, which include the uplink and the downlink transmissions, consider \figurename~\ref{fig:fs})} over the access channel at the SR is defined as
\tcg{
\begin{align}
\rs(\tau) 
          &= \frac{T - \tau - \tp}{T} \e{\eca} {\eca}, \label{eq:rs}
\end{align}}
where $\e{\eca}{\cdot}$ corresponds to an expectation over $\eca$, whose pdf is characterized in Lemma \ref{lm:lm5}. 

\begin{remark} \label{rm:rm3}
\normalfont
\tc{At this point, it is well-known that the performance degradation due to channel estimation in the form of the secondary throughput is inherent to the US. Specifically, the time allocation and the uncertain interference are responsible of this degradation. \tcg{The controlled power, determined in Lemma \ref{lm:lm4}, characterized as a function of the estimation time allows us to regulate the uncertain interference.} As discussed previously in Remark \ref{rm:rm1}, the low estimation time enables a severe control in power, thereby decreasing the secondary throughput. On the other hand, less time resources allocated for the channel estimation increases the secondary throughput. This phenomenon can be captured by observing the variation of the secondary throughput along the estimation time such that the constraints depicted in (\ref{eq:opc}) and (\ref{eq:pc}) are fulfilled. Below, Problem \ref{th:th1} captures this relationship between the estimation time and the secondary throughput defined as an estimation-throughput tradeoff. More importantly, we utilize this tradeoff to determine a suitable estimation time at which the maximum throughput at the SR is achieved.} 
\end{remark}
\begin{theorem} \label{th:th1}
\normalfont
The achievable expected secondary throughput subject to the outage constraint on the uncertain interference and the transmit power constraint at the ST is defined as
\begin{align}
\trs(\ttau) = \maxi_{\tau}  & \text{      } {\rs(\tau)}, 
 \label{eq:sys} \\
\text{s.t.} & \text{ } (\ref{eq:opc}), \text{  } (\ref{eq:pc}), \nonumber 
 \end{align}
where $\trs(\ttau)$ corresponds to optimum throughput at $\ttau$.  
\end{theorem}
\begin{IEEEproof}
The constrained optimization problem is solved by substituting $\preg$ from Lemma \ref{lm:lm3}, determined by applying the outage and the transmit power constraints defined in (\ref{eq:opc}) and (\ref{eq:pc}), in (\ref{eq:rs}). 
The pdf of $\eca$, determined in (\ref{eq:den_C}), is used to evaluate the expectation. Following this, we obtain an expression of the expected secondary throughput as a function of $\tau$\footnote{Please note that $\apt$ and $\bpt$ are also functions of $\tau$, see (\ref{eq:para_pt}).}
\tcg{
\begin{equation}
\rs(\tau) = \frac{T - \tau - \tp}{T} \int\limits_{0}^{\infty} x \dc(x) dx. \label{eq:intrs}
\end{equation}
}
Solving numerically the expression in (\ref{eq:intrs}) yields $\ttau$ and $\trs(\ttau)$. 
\end{IEEEproof}
\begin{coro} \label{cor:cor2}
\normalfont
Problem \ref{th:th1} considers the optimization of the expected secondary throughput for the proposed framework that employ power control and considers the effect of the imperfect channel knowledge. \tc{In accordance to Corollary \ref{cor:cor1}, these USs correspond to the USs that operate in the interference-limited regime $\gamma^* \ge \gamma $.} Besides, it is interesting to compare its performance with those USs that employ channel estimation (as proposed in the paper) and satisfy the outage constraint on the \tcg{uncertain interference}, however, employ no power control, i.e., operate at $\pc$. \tc{With regard to Corollary \ref{cor:cor1}, these systems correspond to the ones operating on the curve $\gamma^*  = \gamma$. For the latter approach, the secondary throughput is obtained by substituting $\preg$ with $\pc$ in (\ref{eq:rs}), where $\tau$ in (\ref{eq:rs}) is determined using Corollary \ref{cor:cor1}. Such a comparison allows us to quantify the performance gain procured by the US \tcg{when power control is employed at the ST.}} 
\end{coro}




\subsection{Random Channel}\label{ssec:ltpa}
Here, our objective is to investigate the performance of the proposed approach, where the interacting channels encounter quasi-static block fading. \tcg{Specifically, the performance of the US is analyzed over multiple frames, where every alternating transmission according to the frame structure presented in \figurename~\ref{fig:fs} observes a different channel.} In this regard, we characterize the channel gains $\gpo$, $\gpt$ and $\gs$ according to Nakagami-$m$ fading model. As a consequence, the power gains $\pgpo$, $\pgpt$ and $\pgs$ follow a Gamma distribution \cite{Goldsmith05}, whose corresponding cdfs are defined as \tc{ 
\begin{align}
\fpgpo(x) = 1 - \Gamma\left(\mpo, \frac{\mpo x}{\bpgpo}\right), \label{eq:dis_pgpo}\\
\fpgpt(x) = 1 - \Gamma\left(\mpt, \frac{\mpt x}{\bpgpt}\right), \label{eq:dis_pgpt}\\  
\fpgs(x) = 1 - \Gamma\left(\ms , \frac{\ms x}{\bpgs}\right), \label{eq:dis_pgs}
\end{align}
where $\mpo$, $\mpt$ and $\ms$ represent the $m$ parameter, whereas $\bpgpo$, $\bpgpt$ and $\bpgs$ are the expected values for channels $\pgpo$, $\pgpt$ and $\pgs$, respectively.} The performance analysis subject to channel fading has been considered by Ghasemi \textit{et al.} \cite{Ghasemi06, Ghasemi07}, \tcg{where the authors in \cite{Ghasemi06, Ghasemi07} evaluated the expected data rate while constraining the interference at the PR.} The influence of channel fading (however, without channel estimation signifying the perfect channel knowledge) has been quantified in terms of the outage constraint on the \tcg{uncertain interference}\footnote{\tcg{In case of the perfect channel knowledge, the uncertainty in the interference is because of channel fading only.}}, is given by 
\begin{align}
	\tc{\maxi_{\preg}} & \text{ } \e{\pgpt, \pgs }{\ca} \label{eq:sys_id_fad}, \\
	\text{s.t.} & \text{ } \tc{\p(\pp = \pgpo \preg \ge \ite)} \le \opc, \label{eq:opc_id_fad} 
\end{align}
where $\e{\pgpt, \pgs} {\cdot}$ corresponds to an expectation with respect to $\pgpt, \pgs$, which are entailed in $\ca$, refer to (\ref{eq:Thr_id}).

Despite the knowledge of the fading model, similar to the ideal model depicted for the deterministic channel (Section \ref{ssec:pd}), the characterization in (\ref{eq:sys_id_fad}) and (\ref{eq:opc_id_fad}) \tcg{assumes the perfect knowledge of the realizations of the power gains $(\pgpo, \pgpt, \pgs)$ for the corresponding channels.} In view of this, we extend our proposed framework to investigate the effect of the random channel (channel fading) on the performance of the US.

In this regard, we first determine the expression of the outage constraint on the uncertain interference  
\begin{align}
\smash[b]{\overbrace{\e{\pgpo}{\smash[b]{\underbrace{\p\left( \left( \frac{\eprcvd - \nps}{\ptran}\right) \preg \ge \ite \right)}_{\text{Channel Estimation}}}}}^{\text{Channel Fading}}} \le \opc, \label{eq:opc_fad} \\[0.5em] \nonumber 
\end{align}
where $\eprcvd$ depends on the underlying value of $\pgpo$, \tcg{which is a random variable}. In contrast to the constraint in (\ref{eq:opc_id_fad}), (\ref{eq:opc_fad}) captures the variations due to the channel estimation $(\p(\cdot)$ determined in terms of $\eprcvd$) and the channel fading ($\e{\pgpo}{\cdot}$). Based on (\ref{eq:opc_fad}) and transmit power constraint defined in (\ref{eq:pc}), we further obtain the expression of the controlled power $(\preg)$ for the case with random channel.

\begin{lemma} \label{lm:lm6}
\normalfont
Subject to the outage constraint on the uncertain interference and the transmit power constraint at the ST, the controlled power at the ST under Nakagami-$m$ fading is given by (see top of next page)
\begin{figure*}
\begin{align}
\preg &=  
\begin{cases} 
\int\limits_{0}^{\infty} \Gamma\left(\frac{\tau \fsam (1 + x \ptran/\nps)^2}{2 + 4 x \ptran/\nps}, \frac{\tau \fsam (1 + x \ptran/\nps)}{\nps (2 + 4 x \ptran/\nps)} \left(\frac{\ite\ptran}{\preg} + \nps \right)\right) d \fpgpo(x) = \opc, & \mbox{for } \preg < \pc \\
\pc, & \mbox{for } \preg \ge \pc
\end{cases}.
\label{eq:preg_fad} 
\end{align}
\hrulefill
\end{figure*}
, where 
$\fpgpo(\cdot)$ is defined in (\ref{eq:dis_pgpo}). 
\end{lemma} 
\begin{IEEEproof}
Since it is complicated to obtain a closed form expression of the integral in (\ref{eq:preg_fad}), we evaluate the controlled power numerically.  
\end{IEEEproof}
\tcg{Similar to analysis performed in Corollary \ref{cor:cor1} for the deterministic channel,} below, we determine the performance bound $(\gamma^*)$ in terms of $\tau$ for the random channel.
To this end, we substitute $\preg$ with $\pc$ in the expression (\ref{eq:preg_fad}) 
\begin{align}
\int\limits_{0}^{\infty} \Gamma\Bigg( \frac{\tau \fsam (1 + x \ptran/\nps)^2}{2 + 4 x \ptran/\nps}, \frac{\tau \fsam (1 + x \ptran/\nps)}{\nps (2 + 4 x \ptran/\nps)} \times \nonumber \\  \left(\frac{\ite\ptran}{\pc} + \nps \right)\Bigg) d \fpgpo(x) &\le \opc. \label{eq:or_fad}
\end{align}
{In order to obtain $\gamma^*$, we evaluate the integral on the left side to obtain function of $\pgpo$ and $\tau$. Since no closed form expression of this function is obtained, we represent this function as}
\begin{align*}
g(\pgpo, \tau \fsam) \le \opc. \nonumber   
\end{align*}
{Substituting $\pgpo = \frac{\gamma^* \nps}{\ptran}$ and replacing with equality, we determine $\gamma^*$ for the random channel as} 
\begin{align*}
g\left(\frac{\gamma^* \nps}{\ptran}, \tau \fsam\right) &= \opc. \nonumber  
\end{align*}
\begin{figure}[!t]
\vspace{\figmara}
%
%
%
\psfrag{s03}[b][b]{\fontsize{8}{12}\fontseries{m}\mathversion{normal}\fontshape{n}\selectfont \color[rgb]{0,0,0}\setlength{\tabcolsep}{0pt}\begin{tabular}{c}$\tau$ [ms]\end{tabular}}%
\psfrag{s04}[t][t]{\fontsize{8}{12}\fontseries{m}\mathversion{normal}\fontshape{n}\selectfont \color[rgb]{0,0,0}\setlength{\tabcolsep}{0pt}\begin{tabular}{c}$\gamma$ [dB]\end{tabular}}%
%
\fontsize{8}{12}\fontseries{m}\mathversion{normal}%
\fontshape{n}\selectfont%
%
\psfrag{x01}[t][t]{-20}%
\psfrag{x02}[t][t]{-19}%
\psfrag{x03}[t][t]{-18}%
\psfrag{x04}[t][t]{-17}%
\psfrag{x05}[t][t]{-16}%
\psfrag{x06}[t][t]{-15}%
\psfrag{x07}[t][t]{-14}%
\psfrag{x08}[t][t]{-13}%
\psfrag{x09}[t][t]{-12}%
\psfrag{x10}[t][t]{-11}%
\psfrag{x11}[t][t]{-10}%
%
\psfrag{v01}[r][r]{0}%
\psfrag{v02}[r][r]{2}%
\psfrag{v03}[r][r]{4}%
\psfrag{v04}[r][r]{6}%
\psfrag{v05}[r][r]{8}%
\psfrag{v06}[r][r]{10}%
\psfrag{v07}[r][r]{12}%
\psfrag{v08}[r][r]{14}%
\psfrag{v09}[r][r]{16}%
\psfrag{v10}[r][r]{18}%
\psfrag{v11}[r][r]{20}%
%
%

\centering
\begin{tikzpicture}[scale=1]
\node[anchor=south west,inner sep=0] (image) at (0,0)
{
\includegraphics[width= \figscale]{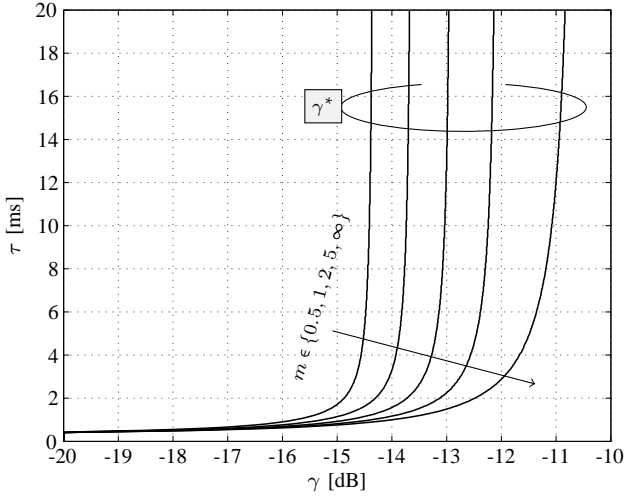}
};
\begin{scope}[x={(image.south east)},y={(image.north west)}]

\draw (0.665,0.82) arc(-250:70:0.2 and 0.05);
\node[draw,fill=gray!10,font=\scriptsize] (text1) at (0.505,0.775) {$\gamma^*$};
\draw[black,->] (0.52,0.31) -- (0.85,0.2);
\node[rotate = 75, font=\scriptsize] at (0.5, 0.38) {$m \in \{0.5, 1,2,5, \infty\}$}; 
\end{scope}
\end{tikzpicture}
\vspace{\figmard}
\caption{An extension of the interference-limited and the power-limited regimes for the US to the random channels, where the channels are subject to Nakagami-$m$ fading. The performance bound ($\gamma^*$) is depicted in terms of estimation time ($\tau$). The different curves demonstrates the severity $(m \in \{0,5,1,2,5, \infty \})$ in fading observed by the channels.}
\label{fig:or_fad}
\vspace{\figmarb}
\end{figure}
\begin{remark} \label{rm:rm2}
\normalfont
\figurename~\ref{fig:or_fad} analyzes the variation of $\gamma^*$ with $\tau$ for different $m \in \{0.5, 1, 2, 5, \infty\}$, where $m = \infty$ represents a deterministic channel. It is observed that $\gamma^*$ attains a lower value as the channel fading becomes severe, thereby enabling the US to operate at low $\gamma$ by extending the interference-limited regime. \tc{Following the analysis from Remark \ref{rm:rm1}, this also reflects that the power control becomes proactive as the severity in the fading increases.} In addition, it is noticed that the deterministic channel is more sensitive to the estimation time as compared to the random channels.  
\end{remark}

After determining the controlled power in Lemma \ref{lm:lm6} that regulates the uncertain interference, we determine the expression of the secondary throughput.
\tcg{\begin{align}
\rs(\tau) &= \e{\eca, \pgpt, \pgs} {\frac{T - \tau - \tp}{T} \eca }, \label{eq:rs_fad} 
\end{align}}
where $\e{\eca, \pgpt, \pgs}{\cdot}$ corresponds to an expectation over $\eca$, $\pgpt$ and $\pgs$, whose cdfs are characterized in Lemma \ref{lm:lm5}, (\ref{eq:dis_pgpt}) and (\ref{eq:dis_pgs}), respectively. It is worth noticing that $\eca$ captures the variations due to channel estimation $\epgs$ and $\eprcvd$,  (\ref{eq:eca}), however, due to channel fading, the underlying values of the channels $\pgpt$ and $\pgs$ are random. In this context, we perform an expectation with respect to $\pgpt$ and $\pgs$, as depicted in (\ref{eq:rs_fad}).
Similar to the deterministic channel, we characterize the estimation-throughput tradeoff for the random channel. 
\begin{theorem} \label{th:th2}
\normalfont
The achievable expected secondary throughput subject to the outage constraint on the uncertain interference and the transmit power constraint at the ST under Nakagami-$m$ fading is defined as
\begin{align}
\trs(\ttau) = \maxi_{\tau}  & \text{      } {\rs(\tau)}, 
 \label{eq:sys_fad} \\
\text{s.t.} & \text{ } (\ref{eq:opc_fad}), \text{  } (\ref{eq:pc}), \nonumber 
 \end{align}
where $\trs(\ttau)$ corresponds to optimum throughput at $\ttau$.  
\end{theorem}

\begin{coro} \label{cor:cor3}
Here, we extend the approach depicted in Corollary \ref{cor:cor2} to study the performance of those USs that employ channel estimation, operates without power control, satisfy the outage constraint and are subjected to Nakagami-$m$ fading. The expected secondary throughput for this particular approach is obtained by replacing $\preg$ in the expression in (\ref{eq:rs_fad}) with $\pc$, where $\tau$ is determined using (\ref{eq:or_fad}). 

\end{coro}
 
\section{Numerical Analysis} \label{sec:num_ana}
In this section, we evaluate the performance of the US based on the proposed model. To accomplish this: (i) we perform simulations to validate the expressions obtained, (ii) we analyze the performance loss incurred due to the channel estimation. In addition, we consider the ideal model to benchmark and evaluate the performance loss. 
Unless stated explicitly, the following values of the parameters are considered for the analysis, $\fsam = \SI{1}{MHz}$, \tcg{$\pgpo$ (deterministic) or $\bpgpo$ (random) $= \SI{-100}{dB}$, $\pgpt$ (deterministic) or $\bpgpt$ (random) = $\SI{-100}{dB}$, $\pgs$ (deterministic) or $\bpgs$ (random) $ = \SI{-80}{dB}$,} $\ite = \SI{-110}{dBm}$, $T = \SI{100}{ms}$, $\opc = 0.10$, $\pc = \SI{0}{dBm}$, $\nps = \SI{-100}{dBm}$, $\gamma = \SI{0}{dB}$, $\ptran = \SI{0}{dBm}$, $\Ks = 10$ and $m \in \{1,5\}$.



\subsection{Deterministic Channel} 

\begin{figure*}[!t]
\vspace{\figmara}
\subfloat[]{
%
%
%
\psfrag{s05}[b][b]{\fontsize{8.5}{12.75}\fontseries{m}\mathversion{normal}\fontshape{n}\selectfont \color[rgb]{0,0,0}\setlength{\tabcolsep}{0pt}\begin{tabular}{c}$\preg$ [dBm]\end{tabular}}%
\psfrag{s06}[t][t]{\fontsize{8.5}{12.75}\fontseries{m}\mathversion{normal}\fontshape{n}\selectfont \color[rgb]{0,0,0}\setlength{\tabcolsep}{0pt}\begin{tabular}{c}$\tau$ [ms]\end{tabular}}%
\psfrag{s10}[][]{\fontsize{10}{15}\fontseries{m}\mathversion{normal}\fontshape{n}\selectfont \color[rgb]{0,0,0}\setlength{\tabcolsep}{0pt}\begin{tabular}{c} \end{tabular}}%
\psfrag{s11}[][]{\fontsize{10}{15}\fontseries{m}\mathversion{normal}\fontshape{n}\selectfont \color[rgb]{0,0,0}\setlength{\tabcolsep}{0pt}\begin{tabular}{c} \end{tabular}}%
\psfrag{s12}[l][l]{\fontsize{8.5}{12.75}\fontseries{m}\mathversion{normal}\fontshape{n}\selectfont \color[rgb]{0,0,0}EM}%
\psfrag{s13}[l][l]{\fontsize{8.5}{12.75}\fontseries{m}\mathversion{normal}\fontshape{n}\selectfont \color[rgb]{0,0,0}IM}%
\psfrag{s14}[l][l]{\fontsize{8.5}{12.75}\fontseries{m}\mathversion{normal}\fontshape{n}\selectfont \color[rgb]{0,0,0}EM}%
%
\fontsize{8.5}{12.75}\fontseries{m}\mathversion{normal}%
\fontshape{n}\selectfont%
%
\psfrag{x01}[t][t]{$10^{-3}$}%
\psfrag{x02}[t][t]{$10^{-2}$}%
\psfrag{x03}[t][t]{$10^{-1}$}%
\psfrag{x04}[t][t]{$10^{0}$}%
\psfrag{x05}[t][t]{$10^{1}$}%
%
\psfrag{v01}[r][r]{-20}%
\psfrag{v02}[r][r]{-19}%
\psfrag{v03}[r][r]{-18}%
\psfrag{v04}[r][r]{-17}%
\psfrag{v05}[r][r]{-16}%
\psfrag{v06}[r][r]{-15}%
\psfrag{v07}[r][r]{-14}%
\psfrag{v08}[r][r]{-13}%
\psfrag{v09}[r][r]{-12}%
\psfrag{v10}[r][r]{-11}%
\psfrag{v11}[r][r]{-10}%
\psfrag{v12}[r][r]{-9}%
%
%

\centering
\begin{tikzpicture}[scale=1]
\node[anchor=south west,inner sep=0] (image) at (0,0)
{
\includegraphics[width= \figscale]{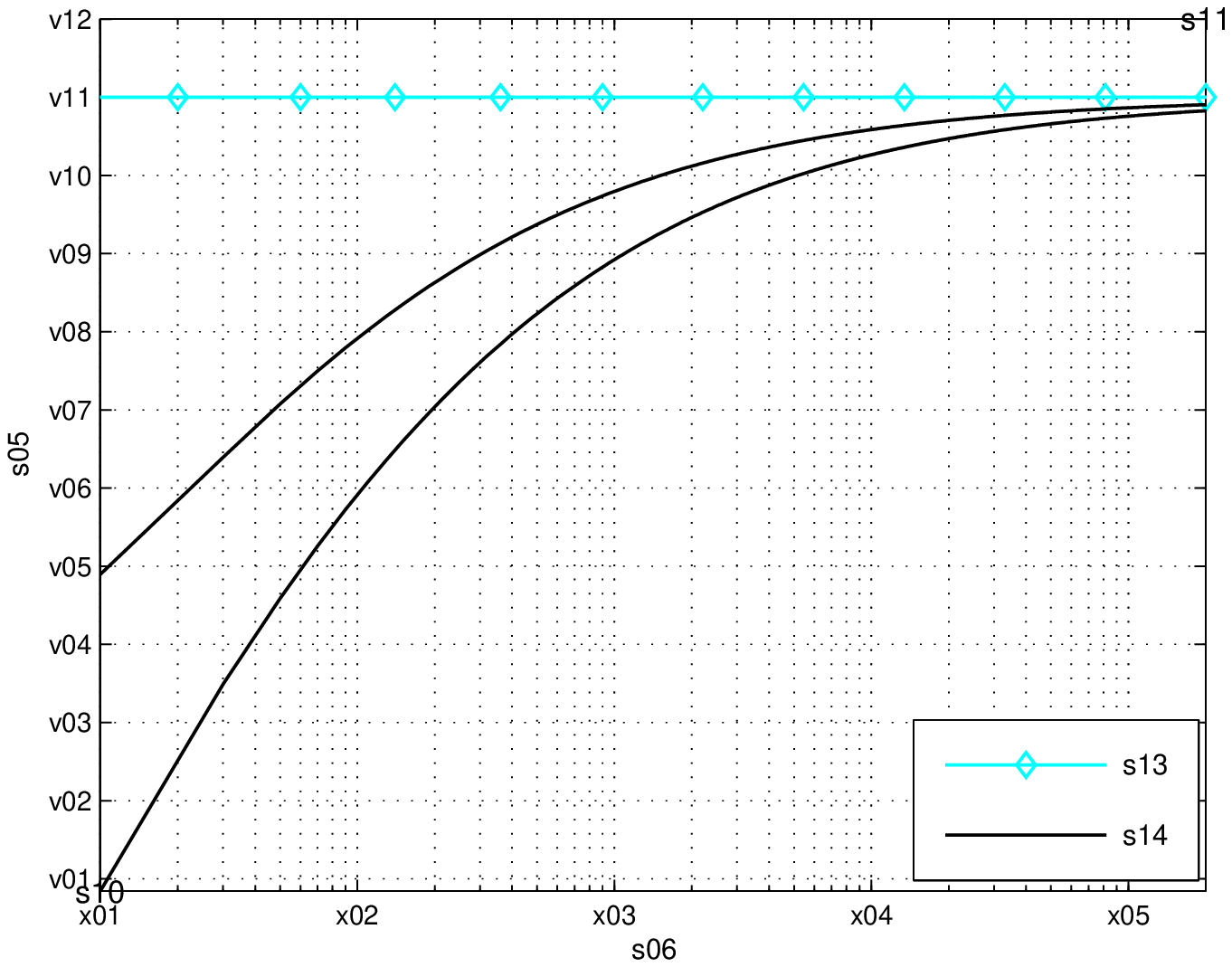}
};
\begin{scope}[x={(image.south east)},y={(image.north west)}]

\draw (0.35,0.56) arc(-130:130:0.007 and 0.021);
\node[draw,fill=gray!10,font=\scriptsize] at (0.44,0.515) {$\opc = 0.01$};

\draw (0.35,0.68) arc(-130:130:0.007 and 0.021);
\node[draw,fill=gray!10,font=\scriptsize] at (0.275,0.758) {$\opc = 0.1$};

\end{scope}
\end{tikzpicture}
\label{fig:Cont_snr}
}
\hfil
\subfloat[]{
%
%
%
\psfrag{s05}[b][b]{\fontsize{8.5}{12.75}\fontseries{m}\mathversion{normal}\fontshape{n}\selectfont \color[rgb]{0,0,0}\setlength{\tabcolsep}{0pt}\begin{tabular}{c}$\rs(\tau)$ [bits/sec/Hz]\end{tabular}}%
\psfrag{s06}[t][t]{\fontsize{8.5}{12.75}\fontseries{m}\mathversion{normal}\fontshape{n}\selectfont \color[rgb]{0,0,0}\setlength{\tabcolsep}{0pt}\begin{tabular}{c}$\tau$ [ms]\end{tabular}}%
\psfrag{s10}[][]{\fontsize{10}{15}\fontseries{m}\mathversion{normal}\fontshape{n}\selectfont \color[rgb]{0,0,0}\setlength{\tabcolsep}{0pt}\begin{tabular}{c} \end{tabular}}%
\psfrag{s11}[][]{\fontsize{10}{15}\fontseries{m}\mathversion{normal}\fontshape{n}\selectfont \color[rgb]{0,0,0}\setlength{\tabcolsep}{0pt}\begin{tabular}{c} \end{tabular}}%
\psfrag{s12}[l][l]{\fontsize{8.5}{12.75}\fontseries{m}\mathversion{normal}\fontshape{n}\selectfont \color[rgb]{0,0,0}Simulated}%
\psfrag{s13}[l][l]{\fontsize{8.5}{12.75}\fontseries{m}\mathversion{normal}\fontshape{n}\selectfont \color[rgb]{0,0,0}IM}%
\psfrag{s14}[l][l]{\fontsize{8.5}{12.75}\fontseries{m}\mathversion{normal}\fontshape{n}\selectfont \color[rgb]{0,0,0}EM}%
\psfrag{s15}[l][l]{\fontsize{8.5}{12.75}\fontseries{m}\mathversion{normal}\fontshape{n}\selectfont \color[rgb]{0,0,0}$\trs(\ttau)$}%
\psfrag{s16}[l][l]{\fontsize{8.5}{12.75}\fontseries{m}\mathversion{normal}\fontshape{n}\selectfont \color[rgb]{0,0,0}Simulated}%
%
\fontsize{8.5}{12.75}\fontseries{m}\mathversion{normal}%
\fontshape{n}\selectfont%
%
\psfrag{x01}[t][t]{0}%
\psfrag{x02}[t][t]{1}%
\psfrag{x03}[t][t]{2}%
\psfrag{x04}[t][t]{3}%
\psfrag{x05}[t][t]{4}%
\psfrag{x06}[t][t]{5}%
\psfrag{x07}[t][t]{6}%
\psfrag{x08}[t][t]{7}%
\psfrag{x09}[t][t]{8}%
\psfrag{x10}[t][t]{9}%
\psfrag{x11}[t][t]{10}%
%
\psfrag{v01}[r][r]{1.6}%
\psfrag{v02}[r][r]{1.8}%
\psfrag{v03}[r][r]{2}%
\psfrag{v04}[r][r]{2.2}%
\psfrag{v05}[r][r]{2.4}%
\psfrag{v06}[r][r]{2.6}%
%
%

\centering
\begin{tikzpicture}[scale=1]
\node[anchor=south west,inner sep=0] (image) at (0,0)
{
\includegraphics[width= \figscale]{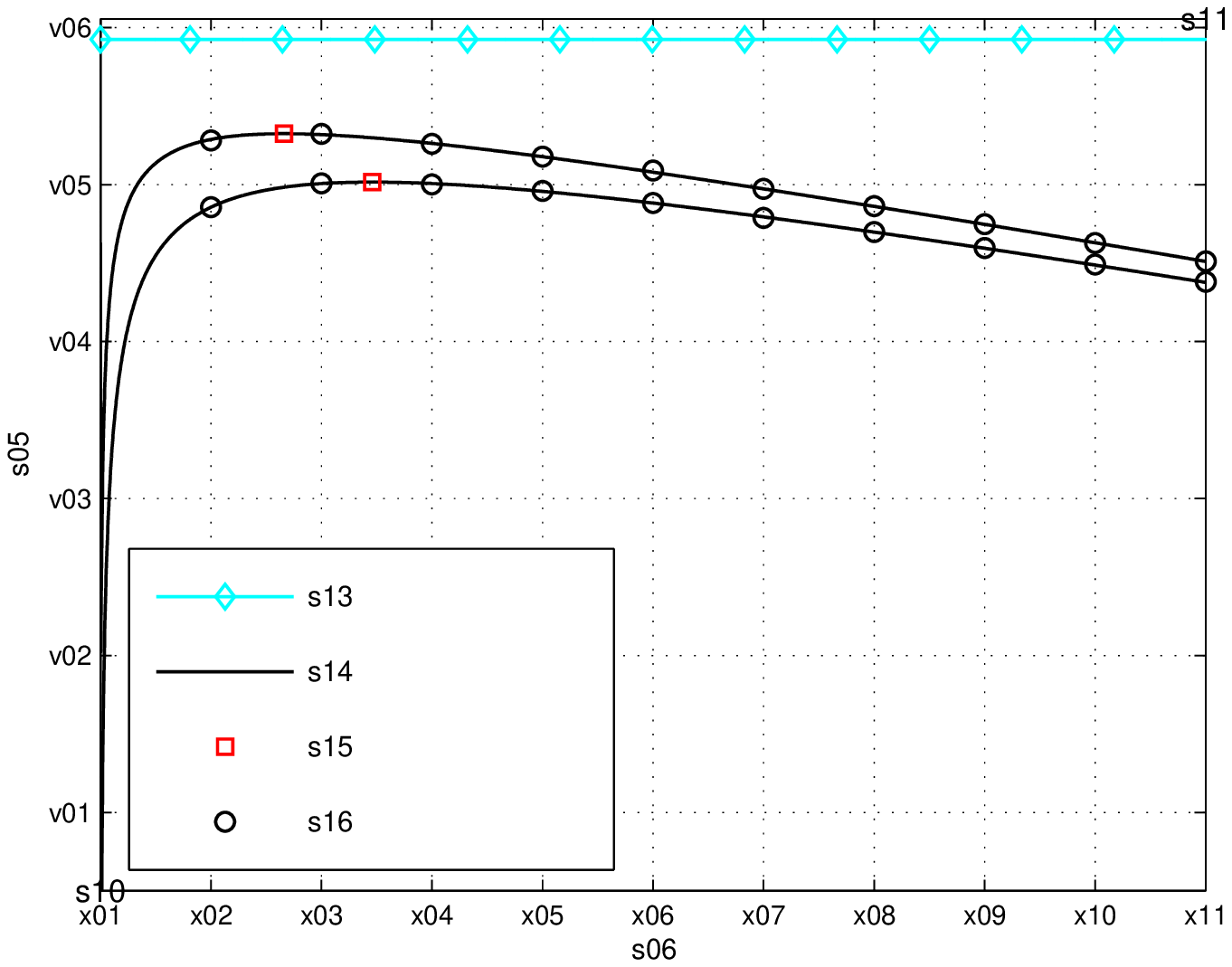}
};
\begin{scope}[x={(image.south east)},y={(image.north west)}]

\draw (0.38,0.78) arc(-130:130:0.005 and 0.015);
\node[draw, fill=gray!10,font=\scriptsize] at (0.39,0.735) {$\opc = 0.01$};

\draw (0.65,0.77) arc(-130:130:0.005 and 0.015);
\node[draw,fill=gray!10,font=\scriptsize] at (0.66,0.838) {$\opc = 0.1$};

\end{scope}
\end{tikzpicture}
\label{fig:ETT}
}
\vspace{\subfigmar}
\caption{(a) Control power versus estimation time with $\gamma = \SI{0}{dB}$, $\opc \in \{0.01, 0.1\}$ and $\pc = \SI{0}{dBm}$. \tcg{(b) Estimation-throughput tradeoff with $\gamma = \SI{0}{dB}$, $\opc \in \{0.01, 0.1\}$ and $\pc = \SI{0}{dBm}$.}}
\vspace{\figmarb}
\end{figure*}
First, we evaluate the performance of the proposed framework in context to the deterministic channel. \figurename~\ref{fig:Cont_snr} considers the variation of $\preg$ versus $\tau$, refer to Corollary \ref{cor:cor1} corresponding to the Ideal Model (IM) and the proposed Estimation Model (EM). \tcg{Since the IM considers the perfect channel knowledge, its controlled power remains invariant to the estimation time.} \tc{In addition, It is noticed that the ST controls its transmit power ($\preg$) more severely for low values of $\tau$, consequently affecting the link budget for the access channel.} 
Besides, \figurename~\ref{fig:ETT} analyzes the performance of the US in terms of the estimation-throughput tradeoff, considered in Problem \ref{th:th1}. 
It can be depicted that the estimation-throughput tradeoff yields a suitable estimation time $\ttau$ that results in an optimum secondary throughput $\trs(\ttau)$. Hereafter, for the performance analysis with respect to the deterministic channel, we consider the theoretical expressions and choose to operate at the suitable estimation time. 

\captionsetup[subfigure]{position=top}
\begin{figure*}[!t]
\vspace{\figmara}
\centering
\subfloat[]{
%
%
%
\psfrag{s05}[b][b]{\fontsize{8.5}{12.75}\fontseries{m}\mathversion{normal}\fontshape{n}\selectfont \color[rgb]{0,0,0}\setlength{\tabcolsep}{0pt}\begin{tabular}{c}$\rs(\ttau)$ [bits/sec/Hz]\end{tabular}}%
\psfrag{s06}[t][t]{\fontsize{8.5}{12.75}\fontseries{m}\mathversion{normal}\fontshape{n}\selectfont \color[rgb]{0,0,0}\setlength{\tabcolsep}{0pt}\begin{tabular}{c}$\gamma$ [dB]\end{tabular}}%
\psfrag{s10}[][]{\fontsize{10}{15}\fontseries{m}\mathversion{normal}\fontshape{n}\selectfont \color[rgb]{0,0,0}\setlength{\tabcolsep}{0pt}\begin{tabular}{c} \end{tabular}}%
\psfrag{s11}[][]{\fontsize{10}{15}\fontseries{m}\mathversion{normal}\fontshape{n}\selectfont \color[rgb]{0,0,0}\setlength{\tabcolsep}{0pt}\begin{tabular}{c} \end{tabular}}%
\psfrag{s12}[l][l]{\fontsize{8.5}{12.75}\fontseries{m}\mathversion{normal}\fontshape{n}\selectfont \color[rgb]{0,0,0}Coro 2}%
\psfrag{s13}[l][l]{\fontsize{8.5}{12.75}\fontseries{m}\mathversion{normal}\fontshape{n}\selectfont \color[rgb]{0,0,0}IM}%
\psfrag{s14}[l][l]{\fontsize{8.5}{12.75}\fontseries{m}\mathversion{normal}\fontshape{n}\selectfont \color[rgb]{0,0,0}EM}%
\psfrag{s15}[l][l]{\fontsize{8.5}{12.75}\fontseries{m}\mathversion{normal}\fontshape{n}\selectfont \color[rgb]{0,0,0}Coro 2}%
%
\fontsize{8.5}{12.75}\fontseries{m}\mathversion{normal}%
\fontshape{n}\selectfont%
%
\psfrag{x01}[t][t]{-20}%
\psfrag{x02}[t][t]{-15}%
\psfrag{x03}[t][t]{-10}%
\psfrag{x04}[t][t]{-5}%
\psfrag{x05}[t][t]{0}%
\psfrag{x06}[t][t]{5}%
\psfrag{x07}[t][t]{10}%
%
\psfrag{v01}[r][r]{0}%
\psfrag{v02}[r][r]{1}%
\psfrag{v03}[r][r]{2}%
\psfrag{v04}[r][r]{3}%
\psfrag{v05}[r][r]{4}%
\psfrag{v06}[r][r]{5}%
\psfrag{v07}[r][r]{6}%
\psfrag{v08}[r][r]{7}%
\psfrag{v09}[r][r]{8}%
\psfrag{v10}[r][r]{9}%
%
%

\centering
\begin{tikzpicture}[scale=1]
\node[anchor=south west,inner sep=0] (image) at (0,0)
{
	\includegraphics[width= \figscale]{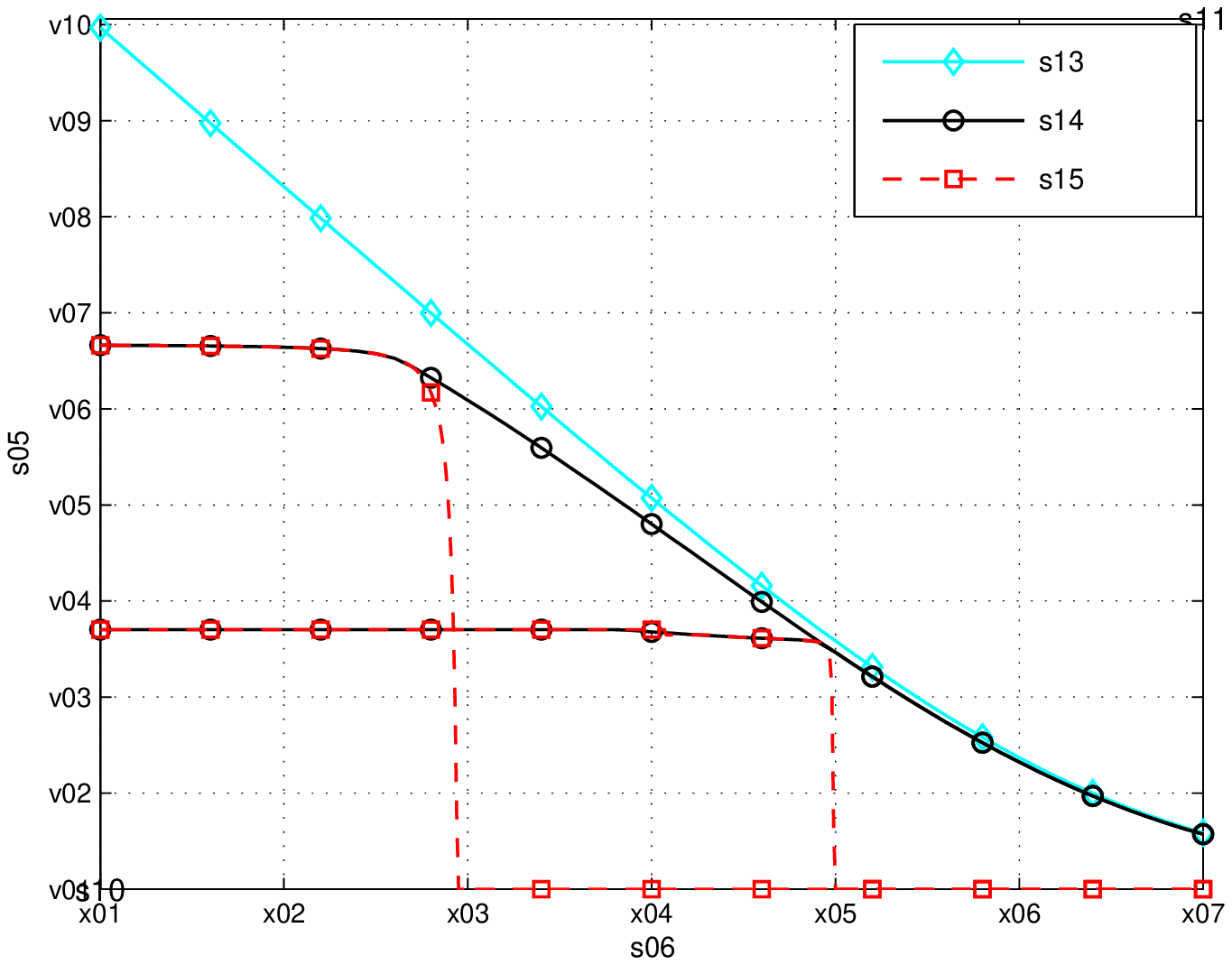}
};
\begin{scope}[x={(image.south east)},y={(image.north west)}]

\draw (0.225,0.615) arc(-160:160:0.01 and 0.03);
\node[draw, fill=gray!10, font=\scriptsize] (text1) at (0.7,0.725) {$\pc = \SI{00}{dBm}$};
\draw (0.275,0.33) arc(-160:160:0.01 and 0.03);
\node[draw, fill=gray!10, font=\scriptsize] (text2) at (0.75,0.44) {$\pc = \SI{-10}{dBm}$};
\draw[black, ->] (text1.west) -- (0.245,0.635);
\draw[black, ->] (text2.west) -- (0.295,0.35);

\end{scope}
\end{tikzpicture}
}
\hfil
\subfloat[]{
%
%
%
\psfrag{s05}[b][b]{\fontsize{8.5}{12.75}\fontseries{m}\mathversion{normal}\fontshape{n}\selectfont \color[rgb]{0,0,0}\setlength{\tabcolsep}{0pt}\begin{tabular}{c}$\rs(\ttau)$ [bits/sec/Hz]\end{tabular}}%
\psfrag{s06}[t][t]{\fontsize{8.5}{12.75}\fontseries{m}\mathversion{normal}\fontshape{n}\selectfont \color[rgb]{0,0,0}\setlength{\tabcolsep}{0pt}\begin{tabular}{c}$\gamma$ [dB]\end{tabular}}%
\psfrag{s10}[][]{\fontsize{10}{15}\fontseries{m}\mathversion{normal}\fontshape{n}\selectfont \color[rgb]{0,0,0}\setlength{\tabcolsep}{0pt}\begin{tabular}{c} \end{tabular}}%
\psfrag{s11}[][]{\fontsize{10}{15}\fontseries{m}\mathversion{normal}\fontshape{n}\selectfont \color[rgb]{0,0,0}\setlength{\tabcolsep}{0pt}\begin{tabular}{c} \end{tabular}}%
\psfrag{s12}[l][l]{\fontsize{8.5}{12.75}\fontseries{m}\mathversion{normal}\fontshape{n}\selectfont \color[rgb]{0,0,0}Coro 2}%
\psfrag{s13}[l][l]{\fontsize{8.5}{12.75}\fontseries{m}\mathversion{normal}\fontshape{n}\selectfont \color[rgb]{0,0,0}IM}%
\psfrag{s14}[l][l]{\fontsize{8.5}{12.75}\fontseries{m}\mathversion{normal}\fontshape{n}\selectfont \color[rgb]{0,0,0}EM}%
\psfrag{s15}[l][l]{\fontsize{8.5}{12.75}\fontseries{m}\mathversion{normal}\fontshape{n}\selectfont \color[rgb]{0,0,0}Coro 2}%
%
\fontsize{8.5}{12.75}\fontseries{m}\mathversion{normal}%
\fontshape{n}\selectfont%
%
\psfrag{x01}[t][t]{-20}%
\psfrag{x02}[t][t]{-15}%
\psfrag{x03}[t][t]{-10}%
\psfrag{x04}[t][t]{-5}%
\psfrag{x05}[t][t]{0}%
\psfrag{x06}[t][t]{5}%
\psfrag{x07}[t][t]{10}%
%
\psfrag{v01}[r][r]{0}%
\psfrag{v02}[r][r]{1}%
\psfrag{v03}[r][r]{2}%
\psfrag{v04}[r][r]{3}%
\psfrag{v05}[r][r]{4}%
\psfrag{v06}[r][r]{5}%
\psfrag{v07}[r][r]{6}%
%
%

\centering
\begin{tikzpicture}[scale=1]
\node[anchor=south west,inner sep=0] (image) at (0,0)
{
	\includegraphics[width= \figscale]{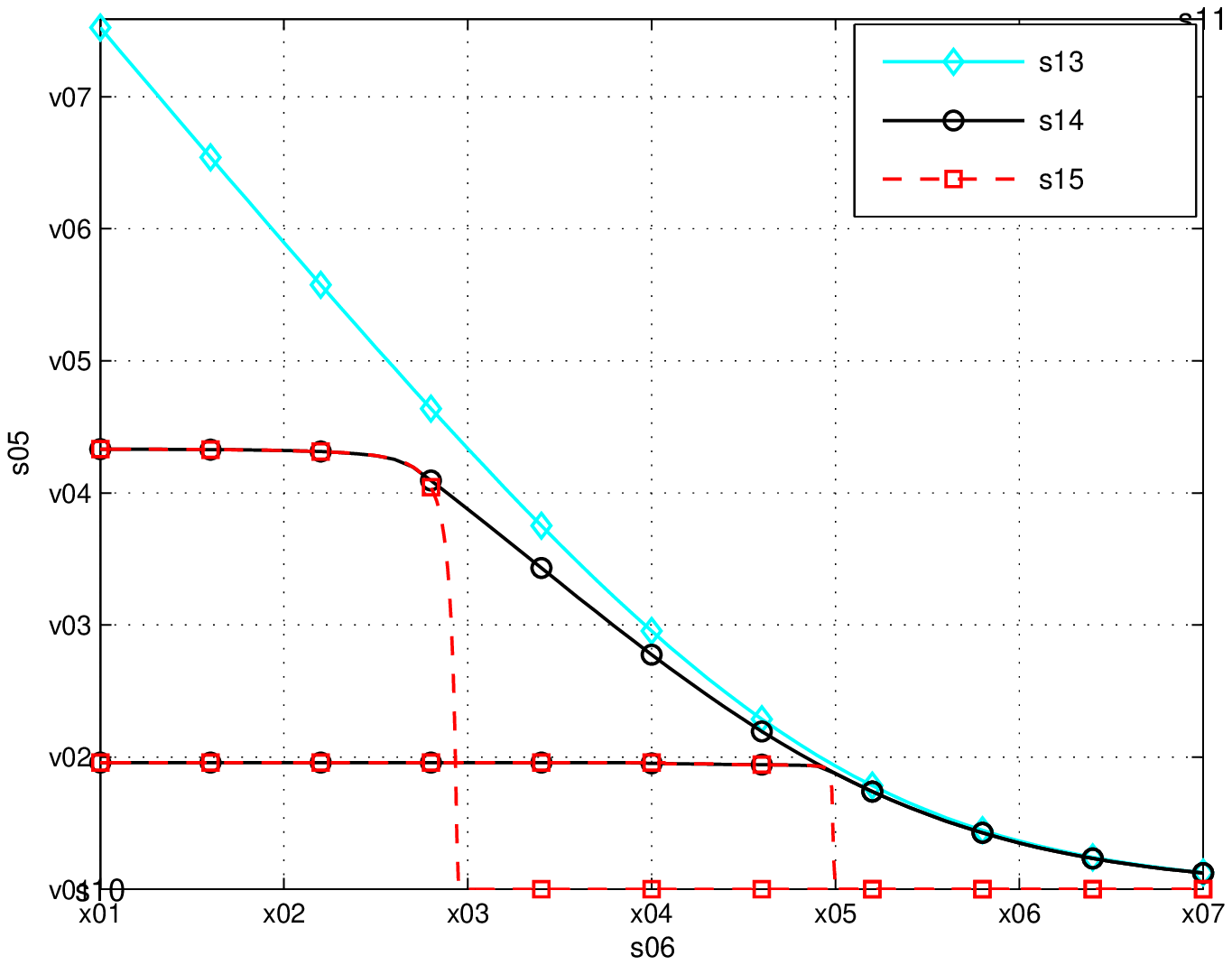}
};
\begin{scope}[x={(image.south east)},y={(image.north west)}]

\draw (0.225,0.51) arc(-160:160:0.01 and 0.03);
\node[draw, fill=gray!10, font=\scriptsize] (text1) at (0.7,0.66) {$\pc = \SI{00}{dBm}$};
\draw (0.275,0.2) arc(-160:160:0.01 and 0.03);
\node[draw, fill=gray!10, font=\scriptsize] (text2) at (0.75,0.35) {$\pc = \SI{-10}{dBm}$};
\draw[black, ->] (text1.west) -- (0.245,0.53);
\draw[black, ->] (text2.west) -- (0.295,0.22);

\end{scope}
\end{tikzpicture}
}
\vspace{\subfigmar}
\caption{\tcg{Optimum secondary throughput $(\rs(\ttau))$ versus the ratio of the received power to noise $(\gamma)$ with $\opc = 0.1$ and $\pc \in \{-10, 0\} \SI{}{dBm}$ for (a) $\pgpt = \SI{-100}{dBm}$ and (b) $\pgpt = \SI{-90}{dBm}$, which translate to an interference power (from the PT) to noise ratio of (a) $\SI{0}{dB}$ and (b) $\SI{10}{dB}$, respectively, at the SR.}}
\label{fig:optT_snr}
\vspace{\figmarb}
\end{figure*}
\tc{To procure further insights, it is necessary to consider the variation of $\trs(\ttau)$ with $\gamma$ for different choices of the \tcg{interference from the PT to noise ratio at the SR over the secondary interference channel} (regulated using $\pgpt \in \{-90, -100\} \SI{}{dBm}$), as depicted in \figurename~\ref{fig:optT_snr}.} \tcg{Due to the limited transmit power at the ST, it is observed that $\trs(\ttau)$ gets saturated below a certain $\gamma$, thereby limiting the performance of the US, depicted in Corollary \ref{cor:cor2}.} Upon increasing $\pc$ from $\SI{-10}{dB}$ to $\SI{0}{dB}$, the point where the saturation is achieved shifts to a lower $\gamma$. This is due to the fact that higher $\pc$ extends the interference-limited regime to a lower $\gamma$. \tcg{In other words, it signifies that, because of low $\gamma$, the secondary system exploits the benefit of operating at a high controlled power.} Particularly for $\pc = \SI{-10}{dBm}$, a severe performance loss indicated by the margin between the IM and the EM is witnessed by the US for $\gamma \le \SI{-2}{dB}$. This concludes that the consideration of the maximum transmit power at the ST is essential while designing the system. Besides this, \figurename~\ref{fig:optT_snr} depicts the performance of the US with no power control, proposed in Corollary \ref{cor:cor2}. As indicated in \figurename~\ref{fig:or}, beyond a certain $\gamma = \gamma^*$, the US with no power control delivers no secondary throughput. In order to \tc{avoid} such situations, the US can exercise power control in order to deliver non-zero secondary throughput. 
\subsection{Random Channel} 
\begin{figure*}[!t]
\vspace{\figmara}
\centering
\subfloat[]{
%
%
%
\psfrag{s05}[b][b]{\fontsize{8.5}{12.75}\fontseries{m}\mathversion{normal}\fontshape{n}\selectfont \color[rgb]{0,0,0}\setlength{\tabcolsep}{0pt}\begin{tabular}{c}$\preg$ [dBm]\end{tabular}}%
\psfrag{s06}[t][t]{\fontsize{8.5}{12.75}\fontseries{m}\mathversion{normal}\fontshape{n}\selectfont \color[rgb]{0,0,0}\setlength{\tabcolsep}{0pt}\begin{tabular}{c}$\tau$ [ms]\end{tabular}}%
\psfrag{s10}[][]{\fontsize{10}{15}\fontseries{m}\mathversion{normal}\fontshape{n}\selectfont \color[rgb]{0,0,0}\setlength{\tabcolsep}{0pt}\begin{tabular}{c} \end{tabular}}%
\psfrag{s11}[][]{\fontsize{10}{15}\fontseries{m}\mathversion{normal}\fontshape{n}\selectfont \color[rgb]{0,0,0}\setlength{\tabcolsep}{0pt}\begin{tabular}{c} \end{tabular}}%
\psfrag{s12}[l][l]{\fontsize{8.5}{12.75}\fontseries{m}\mathversion{normal}\fontshape{n}\selectfont \color[rgb]{0,0,0}Simulated}%
\psfrag{s13}[l][l]{\fontsize{8.5}{12.75}\fontseries{m}\mathversion{normal}\fontshape{n}\selectfont \color[rgb]{0,0,0}IM}%
\psfrag{s14}[l][l]{\fontsize{8.5}{12.75}\fontseries{m}\mathversion{normal}\fontshape{n}\selectfont \color[rgb]{0,0,0}EM}%
\psfrag{s15}[l][l]{\fontsize{8.5}{12.75}\fontseries{m}\mathversion{normal}\fontshape{n}\selectfont \color[rgb]{0,0,0}Simulated}%
%
\fontsize{8.5}{12.75}\fontseries{m}\mathversion{normal}%
\fontshape{n}\selectfont%
%
\psfrag{x01}[t][t]{$10^{-2}$}%
\psfrag{x02}[t][t]{$10^{-1}$}%
\psfrag{x03}[t][t]{$10^{0}$}%
\psfrag{x04}[t][t]{$10^{1}$}%
%
\psfrag{v01}[r][r]{-14}%
\psfrag{v02}[r][r]{-13}%
\psfrag{v03}[r][r]{-12}%
%
%

\centering
\begin{tikzpicture}[scale=1]
\node[anchor=south west,inner sep=0] (image) at (0,0)
{
\includegraphics[width= \figscale]{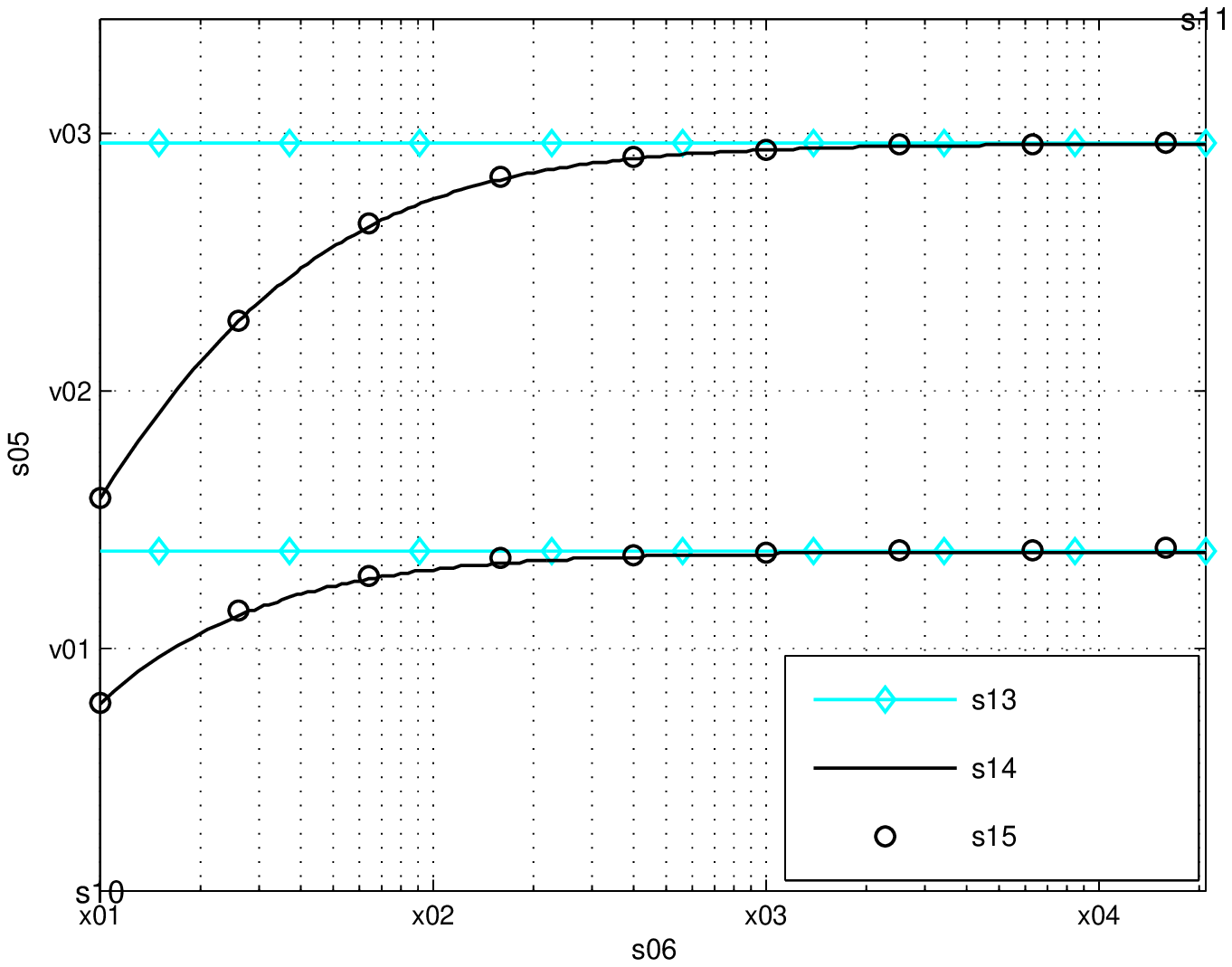}
};
\begin{scope}[x={(image.south east)},y={(image.north west)}]

\draw (0.65,0.405) arc(-130:130:0.007 and 0.021);
\node[draw,fill=gray!10,font=\scriptsize] at (0.66,0.365) {$m = 1$};

\draw (0.65,0.815) arc(-130:130:0.007 and 0.021);
\node[draw,fill=gray!10,font=\scriptsize] at (0.66,0.775) {$m = 5$};



\end{scope}
\end{tikzpicture}
\label{fig:Cont_snr_fad}}
\hfil
\subfloat[]{
%
%
%
\psfrag{s05}[b][b]{\fontsize{8.5}{12.75}\fontseries{m}\mathversion{normal}\fontshape{n}\selectfont \color[rgb]{0,0,0}\setlength{\tabcolsep}{0pt}\begin{tabular}{c}$\rs(\tau)$ [bits/sec/Hz]\end{tabular}}%
\psfrag{s06}[t][t]{\fontsize{8.5}{12.75}\fontseries{m}\mathversion{normal}\fontshape{n}\selectfont \color[rgb]{0,0,0}\setlength{\tabcolsep}{0pt}\begin{tabular}{c}$\tau$ [ms]\end{tabular}}%
\psfrag{s10}[][]{\fontsize{10}{15}\fontseries{m}\mathversion{normal}\fontshape{n}\selectfont \color[rgb]{0,0,0}\setlength{\tabcolsep}{0pt}\begin{tabular}{c} \end{tabular}}%
\psfrag{s11}[][]{\fontsize{10}{15}\fontseries{m}\mathversion{normal}\fontshape{n}\selectfont \color[rgb]{0,0,0}\setlength{\tabcolsep}{0pt}\begin{tabular}{c} \end{tabular}}%
\psfrag{s12}[l][l]{\fontsize{8.5}{12.75}\fontseries{m}\mathversion{normal}\fontshape{n}\selectfont \color[rgb]{0,0,0}Simulated}%
\psfrag{s13}[l][l]{\fontsize{8.5}{12.75}\fontseries{m}\mathversion{normal}\fontshape{n}\selectfont \color[rgb]{0,0,0}IM}%
\psfrag{s14}[l][l]{\fontsize{8.5}{12.75}\fontseries{m}\mathversion{normal}\fontshape{n}\selectfont \color[rgb]{0,0,0}EM}%
\psfrag{s15}[l][l]{\fontsize{8.5}{12.75}\fontseries{m}\mathversion{normal}\fontshape{n}\selectfont \color[rgb]{0,0,0}$\trs(\ttau)$}%
\psfrag{s16}[l][l]{\fontsize{8.5}{12.75}\fontseries{m}\mathversion{normal}\fontshape{n}\selectfont \color[rgb]{0,0,0}Simulated}%
%
\fontsize{8.5}{12.75}\fontseries{m}\mathversion{normal}%
\fontshape{n}\selectfont%
%
\psfrag{x01}[t][t]{$10^{-2}$}%
\psfrag{x02}[t][t]{$10^{-1}$}%
\psfrag{x03}[t][t]{$10^{0}$}%
\psfrag{x04}[t][t]{$10^{1}$}%
%
\psfrag{v01}[r][r]{1}%
\psfrag{v02}[r][r]{1.2}%
\psfrag{v03}[r][r]{1.4}%
\psfrag{v04}[r][r]{1.6}%
\psfrag{v05}[r][r]{1.8}%
\psfrag{v06}[r][r]{2}%
%
%

\begin{tikzpicture}[scale=1]
\node[anchor=south west,inner sep=0] (image) at (0,0)
{
\includegraphics[width= \figscale]{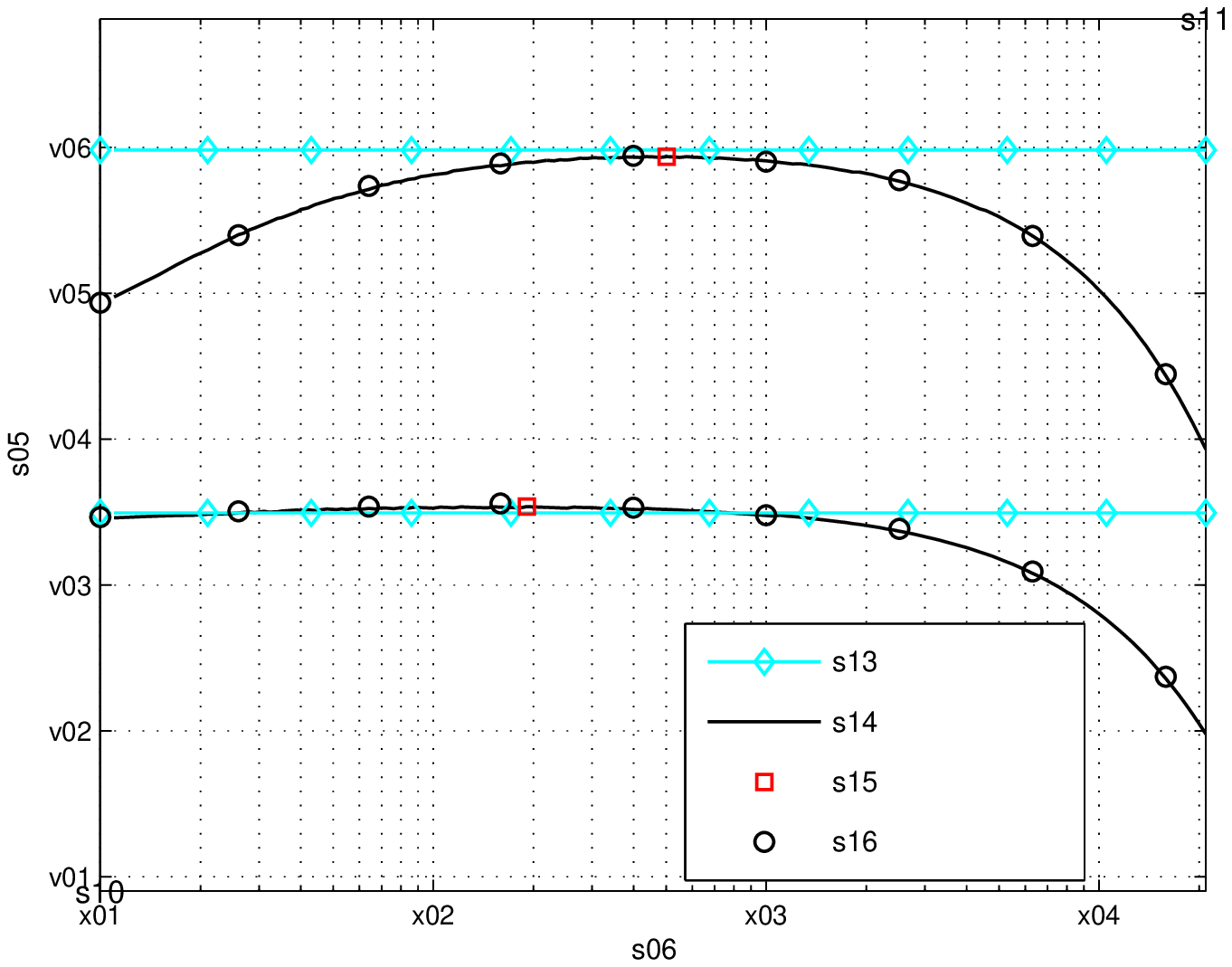}
};
\begin{scope}[x={(image.south east)},y={(image.north west)}]

\draw (0.65,0.445) arc(-130:130:0.007 and 0.021);
\node[draw,fill=gray!10,font=\scriptsize] at (0.66,0.405) {$m = 1$};

\draw (0.65,0.805) arc(-130:130:0.007 and 0.021);
\node[draw,fill=gray!10,font=\scriptsize] at (0.66,0.765) {$m = 5$};

\draw[black,<->] (0.10,0.87) --  node[above = 0.0mm, font=\scriptsize] {Estimation dominant} (0.535,0.87);
\draw[black,<->] (0.54,0.87) --  node[above = 0.0mm, font=\scriptsize] {Channel dominant} (0.95,0.87);

\draw[black,<->] (0.10,0.51) --  node[above = 0.0mm, font=\scriptsize] {Estimation dominant} (0.42,0.51);
\draw[black,<->] (0.425,0.51) --  node[above = 0.0mm, font=\scriptsize] {Channel dominant} (0.95,0.51);

\end{scope}
\end{tikzpicture}
\label{fig:ETT_fad}
}
\vspace{\subfigmar}
\caption{(a) Control power versus estimation time with $\gamma = \SI{0}{dB}$, $\opc = 0.1$ and $\pc = \SI{0}{dBm}$, where Nakagami-$m$ fading is employed to characterize channel fading. \tcg{(b) Estimation-throughput tradeoff with $\gamma = \SI{0}{dB}$, $\opc = 0.1$ and $\pc = \SI{0}{dBm}$ with Nakagami-$m$ fading. The plot classifies the estimation time into the estimation-dominant and the channel-dominant regime.}}
\vspace{\figmarb}
\end{figure*}

Here, we evaluate the performance of the proposed framework, where the interacting channels are under the influence of Nakagami-$m$ fading. For simplification of the analysis, we assume that $m$ is same for all the involved channels, which means $\mpo = \mpt = \ms$. In addition, we investigate the performance under following fading scenarios: (i) severe fading $m=1$, which corresponds to Rayleigh fading\footnote{\tc{Please note that our objective here is to consider the impact of severity in fading on the performance of the US with regard to the channel estimation. The value $m = 1$, which corresponds to a Rayleigh fading, is an obvious representative of a severe fading scenario.}}, and (ii) mild fading $m = 5$. First, we analyze the variation of $\preg$ along the estimation time. It is observed that the mild fading scenario ($m = 5$) is more sensitive to the estimation time, see \figurename~\ref{fig:Cont_snr_fad}. \tcg{In reference to the analysis for the deterministic channel} considered in \figurename~\ref{fig:Cont_snr}, the power control according to the EM saturates with IM at a smaller $\tau$. Complementing the observations carried out in \figurename~\ref{fig:or_fad}, it is concluded that the severe fading scenarios are subjected to a severe power control. 

\tc{Besides, we capture the influence of channel fading on the performance of the US in terms of the estimation-throughput, as depicted in Problem \ref{th:th2}. In this regard, the estimation-throughput tradeoff corresponding to the mild and severe fading scenarios is illustrated in \figurename~\ref{fig:ETT_fad}}. Similar to the case with the deterministic channel, it is depicted that for a suitable choice of the estimation time, the performance of the proposed framework that captures the imperfect channel knowledge is comparable to the ideal conditions in terms of the achievable secondary throughput. Since the US is subjected to the variations from the channel estimation and the channel fading, we classify the estimation time into an estimation-dominant regime and a channel-dominant regime. These regimes signify that the estimation time can only reduce the imperfections (incurred in the US) due to the channel estimation, however, beyond a certain estimation time ($\ttau$), the time resources allocated for channel estimation slightly contribute to the performance improvement (in terms of the controlled power, which finally affects the secondary throughput) and mainly result in the performance degradation (due to the factor \tcg{$\frac{T - \tau- \tp}{T}$} in (\ref{eq:rs_fad})) in the secondary throughput. 
\captionsetup[subfigure]{position=top}
\begin{figure*}[!t]
\vspace{\figmara}
\centering
\subfloat[]{
%
%
%
\psfrag{s05}[b][b]{\fontsize{8.5}{12.75}\fontseries{m}\mathversion{normal}\fontshape{n}\selectfont \color[rgb]{0,0,0}\setlength{\tabcolsep}{0pt}\begin{tabular}{c}$\rs(\ttau)$ [bits/sec/Hz]\end{tabular}}%
\psfrag{s06}[t][t]{\fontsize{8.5}{12.75}\fontseries{m}\mathversion{normal}\fontshape{n}\selectfont \color[rgb]{0,0,0}\setlength{\tabcolsep}{0pt}\begin{tabular}{c}$\gamma$ [dB]\end{tabular}}%
\psfrag{s10}[][]{\fontsize{10}{15}\fontseries{m}\mathversion{normal}\fontshape{n}\selectfont \color[rgb]{0,0,0}\setlength{\tabcolsep}{0pt}\begin{tabular}{c} \end{tabular}}%
\psfrag{s11}[][]{\fontsize{10}{15}\fontseries{m}\mathversion{normal}\fontshape{n}\selectfont \color[rgb]{0,0,0}\setlength{\tabcolsep}{0pt}\begin{tabular}{c} \end{tabular}}%
\psfrag{s12}[l][l]{\fontsize{8.5}{12.75}\fontseries{m}\mathversion{normal}\fontshape{n}\selectfont \color[rgb]{0,0,0}Coro 3}%
\psfrag{s13}[l][l]{\fontsize{8.5}{12.75}\fontseries{m}\mathversion{normal}\fontshape{n}\selectfont \color[rgb]{0,0,0}IM}%
\psfrag{s14}[l][l]{\fontsize{8.5}{12.75}\fontseries{m}\mathversion{normal}\fontshape{n}\selectfont \color[rgb]{0,0,0}EM}%
\psfrag{s15}[l][l]{\fontsize{8.5}{12.75}\fontseries{m}\mathversion{normal}\fontshape{n}\selectfont \color[rgb]{0,0,0}Coro 3}%
%
\fontsize{8.5}{12.75}\fontseries{m}\mathversion{normal}%
\fontshape{n}\selectfont%
%
\psfrag{x01}[t][t]{-20}%
\psfrag{x02}[t][t]{-15}%
\psfrag{x03}[t][t]{-10}%
\psfrag{x04}[t][t]{-5}%
\psfrag{x05}[t][t]{0}%
\psfrag{x06}[t][t]{5}%
\psfrag{x07}[t][t]{10}%
%
\psfrag{v01}[r][r]{0}%
\psfrag{v02}[r][r]{1}%
\psfrag{v03}[r][r]{2}%
\psfrag{v04}[r][r]{3}%
\psfrag{v05}[r][r]{4}%
\psfrag{v06}[r][r]{5}%
\psfrag{v07}[r][r]{6}%
\psfrag{v08}[r][r]{7}%
\psfrag{v09}[r][r]{8}%
%
%

\centering
\begin{tikzpicture}[scale=1]
\node[anchor=south west,inner sep=0] (image) at (0,0)
{
	\includegraphics[width= \figscale]{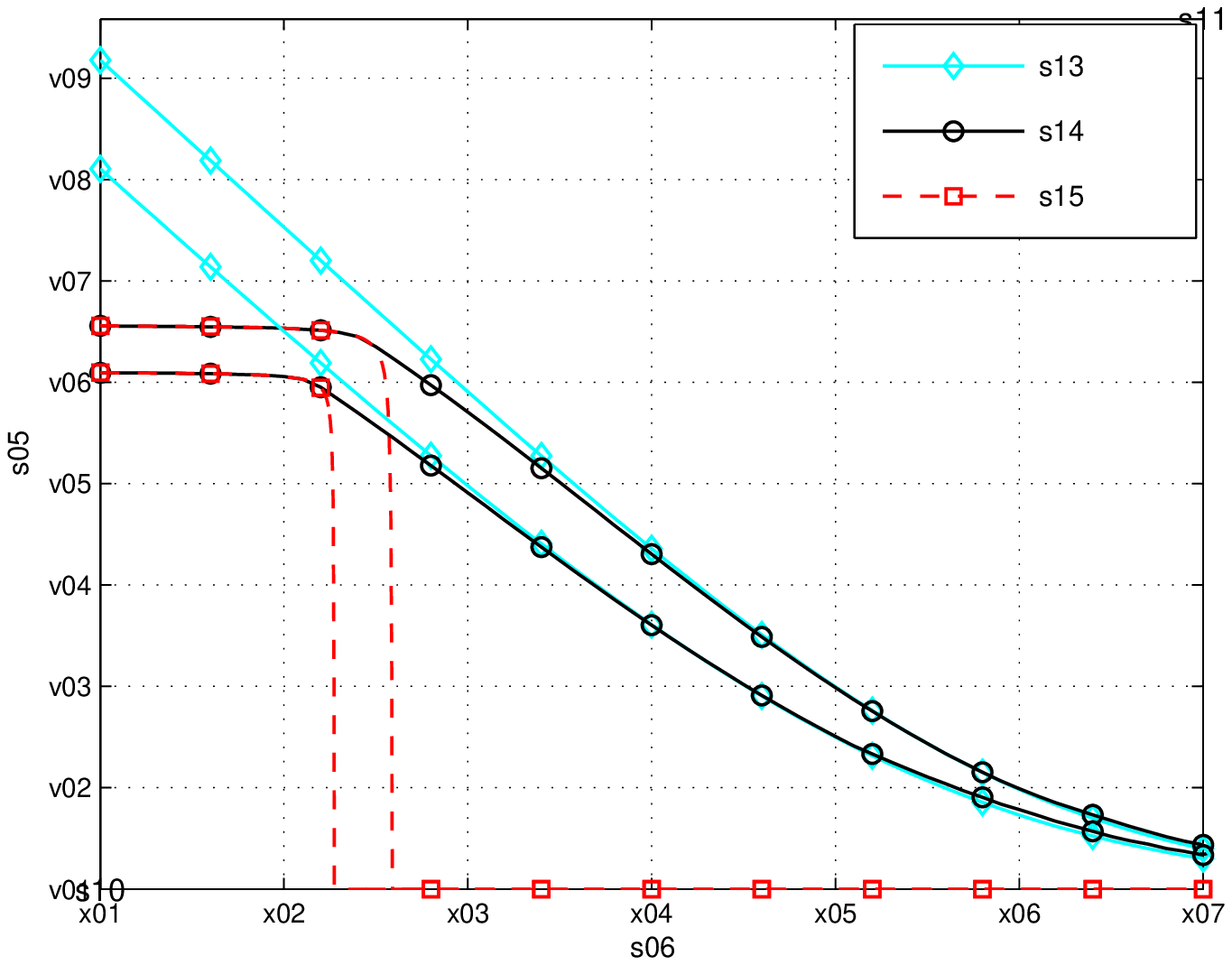}
};
\begin{scope}[x={(image.south east)},y={(image.north west)}]


\draw (0.4,0.54) arc(-160:160:0.007 and 0.021);
\node[draw, fill=gray!10, font=\scriptsize] (text3) at (0.55,0.7) {$m = 5$};
\draw[black, ->] (text3.south) -- (0.412,0.562);

\draw (0.43,0.43) arc(-160:160:0.007 and 0.021);
\node[draw, fill=gray!10, font=\scriptsize] (text4) at (0.58,0.58) {$m = 1$};
\draw[black, ->] (text4.south) -- (0.442,0.452);

\end{scope}
\end{tikzpicture}
}
\hfil
\subfloat[]{
%
%
%
\psfrag{s05}[b][b]{\fontsize{8.5}{12.75}\fontseries{m}\mathversion{normal}\fontshape{n}\selectfont \color[rgb]{0,0,0}\setlength{\tabcolsep}{0pt}\begin{tabular}{c}$\rs(\ttau)$ [bits/sec/Hz]\end{tabular}}%
\psfrag{s06}[t][t]{\fontsize{8.5}{12.75}\fontseries{m}\mathversion{normal}\fontshape{n}\selectfont \color[rgb]{0,0,0}\setlength{\tabcolsep}{0pt}\begin{tabular}{c}$\gamma$ [dB]\end{tabular}}%
\psfrag{s10}[][]{\fontsize{10}{15}\fontseries{m}\mathversion{normal}\fontshape{n}\selectfont \color[rgb]{0,0,0}\setlength{\tabcolsep}{0pt}\begin{tabular}{c} \end{tabular}}%
\psfrag{s11}[][]{\fontsize{10}{15}\fontseries{m}\mathversion{normal}\fontshape{n}\selectfont \color[rgb]{0,0,0}\setlength{\tabcolsep}{0pt}\begin{tabular}{c} \end{tabular}}%
\psfrag{s12}[l][l]{\fontsize{8.5}{12.75}\fontseries{m}\mathversion{normal}\fontshape{n}\selectfont \color[rgb]{0,0,0}Coro 3}%
\psfrag{s13}[l][l]{\fontsize{8.5}{12.75}\fontseries{m}\mathversion{normal}\fontshape{n}\selectfont \color[rgb]{0,0,0}IM}%
\psfrag{s14}[l][l]{\fontsize{8.5}{12.75}\fontseries{m}\mathversion{normal}\fontshape{n}\selectfont \color[rgb]{0,0,0}EM}%
\psfrag{s15}[l][l]{\fontsize{8.5}{12.75}\fontseries{m}\mathversion{normal}\fontshape{n}\selectfont \color[rgb]{0,0,0}Coro 3}%
%
\fontsize{8.5}{12.75}\fontseries{m}\mathversion{normal}%
\fontshape{n}\selectfont%
%
\psfrag{x01}[t][t]{-20}%
\psfrag{x02}[t][t]{-15}%
\psfrag{x03}[t][t]{-10}%
\psfrag{x04}[t][t]{-5}%
\psfrag{x05}[t][t]{0}%
\psfrag{x06}[t][t]{5}%
\psfrag{x07}[t][t]{10}%
%
\psfrag{v01}[r][r]{0}%
\psfrag{v02}[r][r]{1}%
\psfrag{v03}[r][r]{2}%
\psfrag{v04}[r][r]{3}%
\psfrag{v05}[r][r]{4}%
\psfrag{v06}[r][r]{5}%
\psfrag{v07}[r][r]{6}%
%
%

\centering
\begin{tikzpicture}[scale=1]
\node[anchor=south west,inner sep=0] (image) at (0,0)
{
	\includegraphics[width= \figscale]{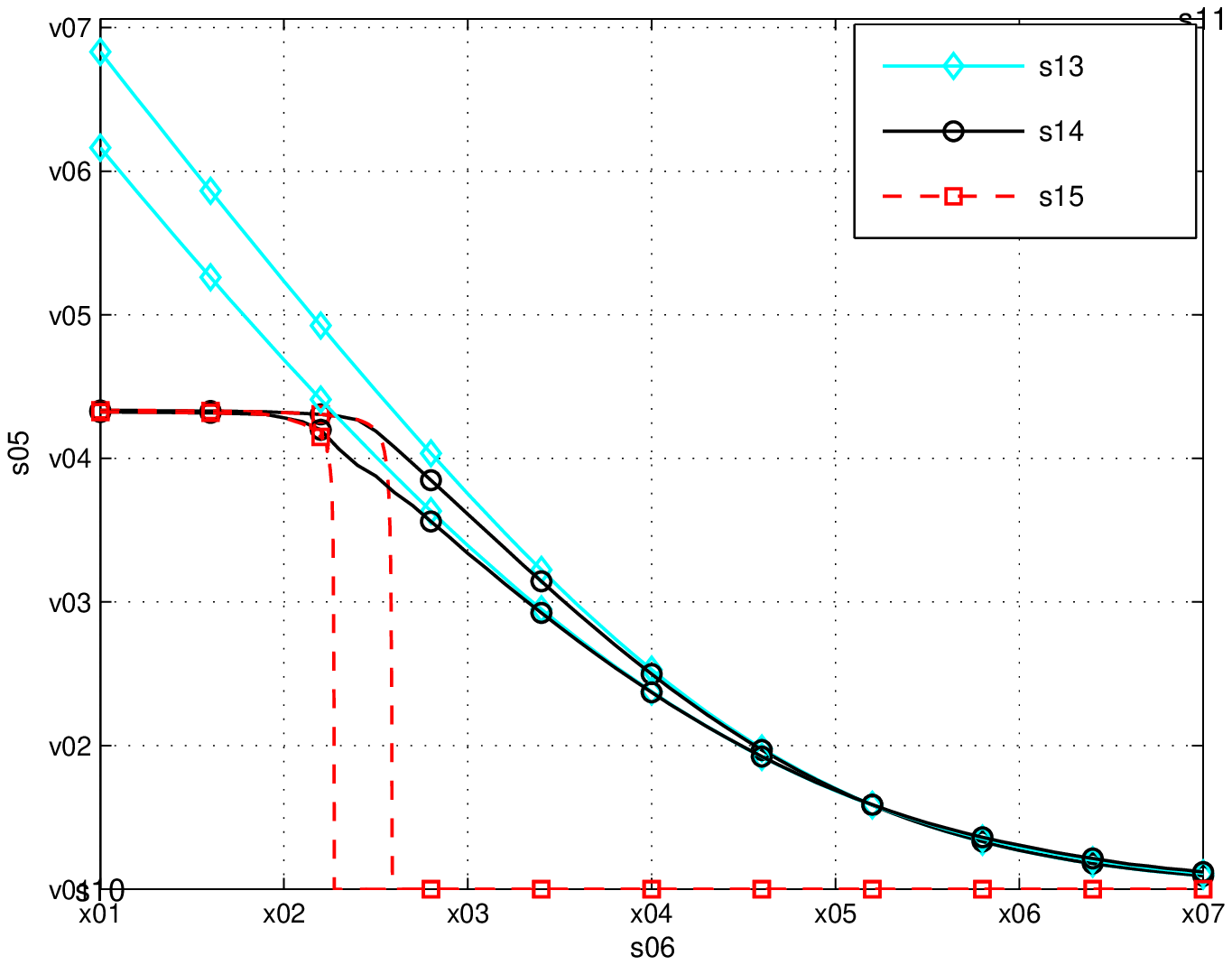}
};
\begin{scope}[x={(image.south east)},y={(image.north west)}]


\draw (0.4,0.435) arc(-160:160:0.007 and 0.021);
\node[draw, fill=gray!10, font=\scriptsize] (text3) at (0.55,0.585) {$m = 5$};
\draw[black, ->] (text3.south) -- (0.412,0.457);

\draw (0.462,0.325) arc(-160:160:0.007 and 0.021);
\node[draw, fill=gray!10, font=\scriptsize] (text4) at (0.38,0.3) {$m = 1$};
\draw[black, ->] (text4.east) -- (0.462,0.32);

\end{scope}
\end{tikzpicture}
}
\vspace{\subfigmar}
\caption{\tcg{Optimum secondary throughput $(\rs(\ttau))$ versus the ratio of the received power to noise $(\gamma)$ for Nakagami-$m$ fading with $\opc = 0.1$ and $\pc = 0$ \SI{}{dBm} for (a) $\pgpt = \SI{-100}{dBm}$ and (b) $\pgpt = \SI{-90}{dBm}$, which correspond to the \tcg{interference} power (from the PT) to noise ratio of (a) $\SI{0}{dB}$ and (b) $\SI{10}{dB}$, respectively, at the SR.}}
\label{fig:optT_snr_fad}
\vspace{\figmarb}
\end{figure*}

Upon determining the optimum secondary throughput $(\rs(\ttau))$ using the estimation-throughput tradeoff, \tcg{we consider the variation of the $\rs(\ttau)$ along the received signal (from the PR) to noise ratio at the ST for different choices of the power gain over the secondary interference channel, which correspond to the \tcg{interference} (from the PT) to noise ratio at the SR, consider \figurename~\ref{fig:optT_snr_fad}.} It is observed that for a large range ($\gamma \ge \SI{-10}{dB}$), the optimum secondary throughput determined by the EM closely follows the secondary throughput depicted by the IM. In addition, \figurename~\ref{fig:optT_snr_fad} considers the performance of the US with no power control,  Corollary \ref{cor:cor3}. Following the discussion in Remark \ref{rm:rm2}, where it was noticed the performance bound ($\gamma^*$) shifts to a lower $\gamma$ when fading becomes severe, \tcg{thus, enabling the ST to carry out a rigorous power control, refer to \figurename~\ref{fig:or_fad}.} This effect is finally translated to the secondary throughput, where $m = 1$ approaches the region with no secondary throughput at a lower $\gamma$ as compared to $m = 5$, consider \figurename~\ref{fig:optT_snr_fad}. 
\section{Conclusion} \label{sec:conc}
\tc{In this paper, we studied the performance of the US from a deployment perspective by putting emphasis on the fact that the knowledge of the interacting channels is pivotal \tcg{to the implementation of the underlay principle over the hardware.}} In this view, a novel approach that incorporates the estimation of the involved channels at the secondary system has been proposed. \tc{Considering the time resources utilized for the channel estimation and the uncertainty due its imperfect knowledge, it has been shown that the channel estimation has a detrimental effect on the performance, leading to its degradation. To tackle the uncertain interference, an outage constraint that precisely regulates the uncertain interference at the PR has been employed. Besides, it has been observed that the operation of the power control at the ST is limited by the maximum transmit power. This limitation, complementing with the channel estimation has been studied in terms of the interference-limited and the power-limited regimes to determine the performance bounds of the US.} 
\tc{Finally, from the perspective of a system designer, an estimation-throughput tradeoff has been established that allows us to determine the achievable secondary throughput for the US. In consideration to the channel fading, it has been observed that the performance degradation is highly prone to the scenarios that are subjected to mild channel fading.} 

In our future work, we plan to extend the proposed analysis to capture the influence of multiple primary and secondary users present in the network on the performance of the US. \tc{In addition, the performance evaluation presented in the paper considers symmetric fading, i.e., the channel gains are subjected to the same value of $m$. However, depending on the deployment scenario, the derived expressions can be utilized to realize asymmetric fading by substituting different values of $m$ corresponding to  the channels. In this regard, we plan to extend the proposed framework to study the influence of asymmetric fading on the performance of the US.} 
\appendix[Proof of Lemma 5] \label{ap:one}
\begin{IEEEproof}
\tc{For simplification, we deal $\frac{\epgs \preg}{\eprcvdsr}$ as individual terms $\epgs \preg$ and $\eprcvdsr$ and determine the pdfs $f_{\epgs \preg}(\cdot)$ and $f_{\eprcvdsr}(\cdot)$ separately.}
Using (\ref{eq:fehs}) in Lemma \ref{lm:lm2}, the pdf of $f_{\epgs \ptran}$ is determined as
\begin{align}
f_{\epgs \preg}(x) =& \frac{1}{\Gamma(\as) (\bs \preg)^{\as}} x^{\as - 1} \times \nonumber \\ & \exp\left( - \frac{x}{ \bs \preg} \right), \label{eq:step1} 
\end{align}
where $\as$ and $\bs$ are defined in (\ref{eq:para_s}).
Similarly, using Lemma \ref{lm:lm3}, the pdf of $\eprcvdsr$ is characterized as
\begin{align}
f_{\eprcvdsr}(x) = \frac{1}{\Gamma(\apt) (\bpt)^{\apt}} x^{\apt - 1} \exp\left( - \frac{x}{ \bpt} \right), \label{eq:step2} 
\end{align}
where $\apt$ and $\bpt$ are defined in (\ref{eq:para_pt}).

Using (\ref{eq:step1}) and (\ref{eq:step2}), we apply Mellin transform \cite{NIST} to determine the pdf of $\frac{\epgs \preg}{\eprcvdsr}$ as
\begin{align}
f_{\frac{\epgs \preg}{\eprcvdsr}}(x) =& \frac{(x)^{\as - 1} \Gamma(\as + \apt)}{\Gamma(\as) \Gamma(\apt) (\bs \preg) ^{\as} \bpt^{\apt}} \times \nonumber \\ &  \left(\frac{1}{\bpt} + \frac{x - 1}{\bs \preg}\right).
\end{align}
Finally, substituting the expression $\frac{\epgs \preg}{\eprcvdsr}$ in $\eca$ yields (\ref{eq:den_C}).
\end{IEEEproof}

\bibliographystyle{IEEEtran}
\bibliography{IEEEabrv,refs}


\end{document}